\documentclass[
notoc,nohyper]{JHEP3} 

\usepackage{epsfig}

\newcommand{\Gammait}{{\mit\Gamma}}

\newcommand{\Pf}{\mathop{\rm Pf}\nolimits}

\newcommand{\U}{\mathop{\rm {}U}}
\newcommand{\re}{\mathop{\rm Re}\nolimits}
\newcommand{\im}{\mathop{\rm Im}\nolimits}
\newcommand{\rmd}{{\rm d}}

\newcommand{\ring}{\mathaccent"7017 }

\newcommand\fverb{\setbox\pippobox=\hbox\bgroup\verb}
\newcommand\fverbdo{\egroup\medskip\noindent%
                        \fbox{\unhbox\pippobox}\ }
\newcommand\fverbit{\egroup\item[\fbox{\unhbox\pippobox}]}
\newbox\pippobox

\title{
A local formulation of lattice Wess-Zumino model with exact $\U(1)_R$ symmetry}

\author{Yoshio KIkukawa\\
Department of Physics, Nagoya University, Nagoya 464-8602,
Japan\\
E-mail: \email{kikukawa@eken.phys.nagoya-u.ac.jp}}
\author{Hiroshi Suzuki\\
Institute of Applied Beam Science, Ibaraki University, Mito 310-8512, Japan\\
E-mail: \email{hsuzuki@mx.ibaraki.ac.jp}}

\received{\today}               
\accepted{\today}               

\preprint{DPNU-04-24\\IU-MSTP/64\\\heplat{0412042}}

\abstract{A lattice Wess-Zumino model is formulated on the basis of
Ginsparg-Wilson fermions. In perturbation theory, our formulation is equivalent
to the formulation by Fujikawa and~Ishibashi and by Fujikawa. Our formulation
is, however, free from a singular nature of the latter formulation due to an
additional auxiliary chiral supermultiplet on a lattice. The model posssesses
an exact $\U(1)_R$ symmetry as a supersymmetric counterpart of the L\"uscher
lattice chiral $\U(1)$ symmetry. A restration of the supersymmetric
Ward-Takahashi identity in the continuum limit is analyzed in renormalized
perturbation theory. In the one-loop level, a supersymmetric continuum limit is
ensured by suitably adjusting a coefficient of a single local
term~$\tilde F^*\tilde F$. The non-renormalization theorem holds to this order
of perturbation theory. In higher orders, on the other hand, coefficents of
local terms with dimension~$\leq4$ that are consistent with the $\U(1)_R$
symmetry have to be adjusted for a supersymmetric continuum limit. The origin
of this complexicity in higher-order loops is clarified on the basis of the
Reisz power counting theorem. Therefore, from a view point of supersymmetry,
the present formulation is not quite better than a lattice Wess-Zumino model
formulated by using Wilson fermions, although a number of coefficients which
require adjustment is much less due to the exact $\U(1)_R$ symmetry. We also
comment on an exact non-linear fermionic symmetry which corresponds to the one
studied by Bonini and Feo; an existence of this exact symmetry itself does not
imply a restoration of supersymmetry in the continuum limit without any
adjustment of parameters.}

\begin{document}

\maketitle 

\section{Introduction}
There has been a renewed interest on non-perturbative formulation of
supersymmetric theories via a spacetime lattice~\cite{Dondi:1976tx}--%
\cite{Aoyama:1998in} in these several years~\cite{Catterall:2001fr}--%
\cite{Bonini:2004pm} (for a recent review with a complete list of references,
see ref.~\cite{Feo:2002yi}). One major idea in these recent studies is to keep
a part of the supersymmetry algebra manifest and infer that this exact symmetry
is strong enough to ensure a fully supersymmetric continuum limit without any
(or with a small number of) adjustment of
parameters~\cite{Catterall:2001fr,Kaplan:2002wv,Kikukawa:2002as,Sugino:2003yb}.
This general strategy, which is also common to some of past
attempts~\cite{Elitzur:1982vh}, has achieved fair success, typically for lower
dimensional supersymmetric theories with an extended supersymmetry (besides a
potential problem of positivity of the measure). An extended supersymmetry
allows a sub-algebra which is consistent with a lattice construction and, due
to the lower-ness of dimensionality, the number of relevant operators, which
potentially violate supersymmetry in the continuum limit, is small. So often a
supersymmetric continuum limit is achieved without any adjustment of
parameters.

Another kind of approaches is to abandon a manifest supersymmetry of a lattice
model from the onset and achieve a supersymmetric continuum limit by adjusting
parameters in the model. This is an approach advocated in
refs.~\cite{Bartels:1982ue,DiVecchia:1983ax,Curci:1986sm,Golterman:1988ta} and
this has been, in our opinion, only a realistic approach to date for $N=1$
supersymmetric theories in 4~dimensions. Here, again, some exact global
symmetries on a lattice can be useful~\cite{Nishimura:1997vg,Aoyama:1998in} to
reduce the number of parameters which require adjustment. For numerical
simulations along this kind of approaches, see
ref.~\cite{Montvay:2001aj}.\footnote{There exist more ambitious approaches
which aim an exact full supersymmetry on a
lattice~\cite{Bartels:1983wm,Itoh:2001rx,D'Adda:2004jb}. Another interesting
observation is that a supersymmetric continuum theory is automatically restored
if a convergence behavior of Feynman integrals in a lattice model is moderate
enough~\cite{Fujikawa:2002ic,Fujikawa:2002pa}.}

In this paper, we adopt the latter attitude and study a lattice formulation of
the 4~dimensional $N=1$ supersymmetric Wess-Zumino model~\cite{Wess:1974tw}.
The Wess-Zumino model is asymptotic non-free and thus the continuum limit, as
a fundamental theory, is expected to be trivial. Nevertheless, it is meaningful
to consider the model as an effective theory with an ultraviolet cutoff. In a
sense, this model is more difficult to formulate on a lattice than
supersymmetric Yang-Mills theories because a quadratic divergence in mass terms
of scalar fields is expected to be prohibited only with presence of an exact
supersymmetry. It is thus deserve to study in its own right. We formulate the
model on the basis of Ginsparg-Wilson
fermions~\cite{Ginsparg:1982bj}--\cite{Luscher:1998pq}. This kind of
formulation has been pursued by Fujikawa and Ishibashi and by
Fujikawa~\cite{Fujikawa:2001ka,Fujikawa:2001ns,Fujikawa:2002ic}. In our
notation, their formulation is expressed as
\begin{eqnarray}
   S&=&a^4\sum_x\Bigl\{{1\over2}\chi^TC(1-{1\over2}aD_2)^{-1}D_1\chi
   -{2\over a}\phi^*D_2\phi+F^*(1-{1\over2}aD_2)^{-1}F
\nonumber\\
   &&\qquad\qquad
   +{1\over2}m\chi^TCP_+\chi+{1\over2}m^*\chi^TCP_-\chi
   +mF\phi+m^*F^*\phi^*
\nonumber\\
   &&\qquad\qquad
   +g\chi^TC\phi P_+\chi+g F\phi^2+g^*\chi^TC\phi^*P_-\chi
   +g^* F^*\phi^{*2}\Bigr\},
\label{onexone}
\end{eqnarray}
where $(\chi,\phi,F)$ denote the chiral multiplet of the Wess-Zumino model
and $D_1$ and $D_2$ are lattice difference operators which will be defined
below. In ref.~\cite{Fujikawa:2001ka}, explicit perturbative calculations in
one-loop order were carried out and it was found that, in the one-loop level,
effects of supersymmetry breaking in the model appear only in wave
function renormalization factors of the chiral multiplet, thus the violation of
supersymmetry is rather moderate in the one-loop level.

One can carry out perturbative calculations on the basis of the
action~(\ref{onexone}) without any problem. However, as pointed out in
ref.~\cite{Fujikawa:2001ns} (see also ref.~\cite{Fujikawa:2002is}), the action 
itself is singular because the operator~$1-(1/2)aD_2$ always has zero modes.
This also implies that the kinetic operator in eq.~(\ref{onexone}) is
non-local. Thus the meaning of the model in a non-perturbative level is not
clear.

In this paper, we first formulate a {\it non-singular\/} local lattice action
for the Wess-Zumino model which is, in perturbation theory, equivalent to the
formulation based on the action~(\ref{onexone}). This is achieved by
introducing a non-dynamical auxiliary chiral multiplet on a lattice which
decouples in the continuum limit. Due to the Ginsparg-Wilson relation, when
$m=0$, our model possesses a lattice analogue of the $\U(1)_R$ symmetry which
is supersymmetric counterpart of the L\"uscher lattice chiral $\U(1)$
symmetry~\cite{Luscher:1998pq}.\footnote{In the action~(\ref{onexone}), this
$\U(1)_R$ symmetry is trivially realized as
$\delta_\alpha\chi=+i\alpha\gamma_5\chi$,
$\delta_\alpha\phi=-2i\alpha\phi$ and~$\delta_\alpha F=+4i\alpha F$. See
section~2.} This is the symmetry that was pointed out for a free theory in
ref.~\cite{Aoyama:1998in}. These are contents of section~2.

Next, we study a restoration of supersymmetry in the continuum limit by using a
lattice version of the Ward-Takahashi identity. We carry out an explicit
one-loop evaluation of the supersymmetry breaking term in the Ward-Takahashi
identity and observe that effects of supersymmetry breaking in the present
model appear only in the wave function renormalization of the auxiliary
field~$\tilde F$ in the continuum limit (in the one-loop level). We then
present a general argument for higher loop contributions of supersymmetry
breaking in renormalized perturbation theory. Unfortunately, the general
power counting argument which is based on the Reisz
theorem~\cite{Reisz:1987da,Luscher:1988sd} indicates that all supersymmetric
non-invariant local terms with the mass dimension~$\leq4$ are radiatively
induced, unless a term is forbidden by the $\U(1)_R$ global symmetry that is
manifest in our formulation. We clarify the reason why the one-loop result is
so simple and higher loop corrections are expected to be destructive. In terms
of the power counting theorem, the supersymmetry breaking term, which is a
consequence of a violation of the Leibniz rule on the lattice, behaves as a
non-derivative coupling in one-loop diagrams while it behaves as a derivative
coupling in higher loop diagrams. This peculiar behavior of the supersymmetry
breaking term makes the situation in higher loop diagrams involved. As a
conclusion, from a view of supersymmetry restoration, our formulation is not
quite better than the formulation based on the Wilson
fermion~\cite{Bartels:1982ue}, although some of super non-invariant local terms
are prohibited by the exact $\U(1)_R$ symmetry. (Sec.~3)

In the final part of this paper, we will comment on an exact non-linear
fermionic symmetry in our formulation which corresponds to the symmetry
recently studied in~ref.~\cite{Bonini:2004pm} in the context of the
Fujikawa-Ishibashi formulation. This symmetry is nothing but the ``lattice
supersymmetry'' utilized in ref.~\cite{Golterman:1988ta} for 2~dimensional
Wess-Zumino model. As noted in ref.~\cite{Golterman:1988ta} and as indicated
from the results of ref.~\cite{Fujikawa:2001ka} and of ours, a presence of this
symmetry itself does not imply an automatic restoration of supersymmetry in the
continuum limit without any adjustment of parameters (although the non-linear
symmetry reduces to the standard supersymmetry in the classical continuum
limit). We clarify this point.

Throughout this paper, the lattice spacing will be denoted by~$a$.

\section{The model}
\subsection{Action and its symmetries}
Our staring point is the chiral invariant lattice Yukawa model of
ref.~\cite{Luscher:1998pq}:
\begin{eqnarray}
   S&=&a^4\sum_x\Bigl\{\overline\psi D\psi
   -{2\over a}\overline\Psi\Psi
   +2g(\overline\psi+\overline\Psi)
   \Bigl(\phi+{m\over2g}\Bigr)P_+(\psi+\Psi)
\nonumber\\
   &&\qquad\qquad\qquad\qquad\qquad
   +2g^*(\overline\psi+\overline\Psi)
   \Bigl(\phi^*+{m^*\over2g^*}\Bigr)P_-(\psi+\Psi)\Bigr\},
\label{twoxone}
\end{eqnarray}
where the field~$\Psi$ is a non-dynamical auxiliary fermionic field and
$P_\pm=(1\pm\gamma_5)/2$. In this expression, we have shifted the scalar field
as $\phi\to\phi+m/(2g)$ to generate mass terms for fermions. As the lattice
Dirac operator~$D$, we adopt the overlap-Dirac
operator~\cite{Kikukawa:1997qh,Neuberger:1998fp} defined
by\footnote{$\partial_\mu f(x)=\{f(x+a\hat\mu)-f(x)\}/a$
and~$\partial_\mu^*f(x)=\{f(x)-f(x-a\hat\mu)\}/a$ are the forward and backward
difference operators, respectively.}
\begin{equation}
   D={1\over2}\{1-A(A^\dagger A)^{-1/2}\},\qquad
   A=1-aD_{\rm w},\qquad
   D_{\rm w}={1\over2}\{\gamma_\mu(\partial_\mu^*+\partial_\mu)
   -a\partial_\mu^*\partial_\mu\},
\end{equation}
which obeys the Ginsparg-Wilson
relation~$\gamma_5D+D\gamma_5=aD\gamma_5D$~\cite{Ginsparg:1982bj}. Thanks to
this relation, the action with~$m=0$ is invariant under a lattice chiral
transformation of the following form~\cite{Luscher:1998pq}
\begin{eqnarray}
   &&\delta_\alpha\psi
   =i\alpha\gamma_5(1-{1\over2}aD)\psi+i\alpha\gamma_5\Psi,\qquad
   \delta_\alpha\Psi=i\alpha\gamma_5{1\over2}aD\psi,
\nonumber\\
   &&\delta_\alpha\overline\psi
   =i\alpha\overline\psi(1-{1\over2}aD)\gamma_5
   +i\alpha\overline\Psi\gamma_5,
   \qquad
   \delta_\alpha\overline\Psi=i\alpha\overline\psi{1\over2}aD\gamma_5,
\nonumber\\
   &&\delta_\alpha\phi=-2i\alpha\phi,\qquad
   \delta_\alpha\phi^*=2i\alpha\phi^*,
\end{eqnarray}
where $\alpha$ is an infinitesimal real parameter. This transformation is
designed so that a sum of fields, say $\psi+\Psi$, transforms in a standard
way, $\delta_\alpha(\psi+\Psi)=i\alpha\gamma_5(\psi+\Psi)$. Thus a breaking of
this chiral symmetry due to the presence of mass terms has a simple structure
as in the continuum theory. The auxiliary fermion~$\Psi$ is introduced to make
a chiral transformation of this standard form and the Ginsparg-Wilson relation
(which implies a non-standard chiral property of the lattice Dirac
operator~$D$) compatible. 

To define the Wess-Zumino model, we need to reduce degrees of freedom of the
Dirac field~$\psi$ to the Majorana one. Since the chiral
projectors~$P_\pm=(1\pm\gamma_5)/2$ in the Yukawa interaction term are ordinary
ones, the Majorana reduction (see ref.~\cite{Fujikawa:2001ka}) can be applied
straightforwardly. Namely, we make substitutions\footnote{$C$ is the
charge conjugation matrix which satisfies $C\gamma_\mu C^{-1}=-\gamma_\mu^T$,
$C\gamma_5 C^{-1}=\gamma_5^T$, $C^\dagger C=1$ and~$C^T=-C$.}
\begin{eqnarray}
   &&\psi=(\chi+i\eta)/\sqrt{2},\qquad
   \overline\psi=(\chi^TC-i\eta^TC)/\sqrt{2},
\nonumber\\
   &&\Psi=(X+iY)/\sqrt{2},\qquad\overline\Psi=(X^TC-iY^TC)/\sqrt{2},
\end{eqnarray}
in the action. Noting relations
\begin{equation}
   (CD)^T=-CD,\qquad(CP_\pm)=-CP_\pm,
\end{equation}
we find that the action decomposes into two independent systems. By taking only
terms including $\chi$ and $X$, we have\footnote{This Majorana reduction, in a
level of the functional integral, corresponds to the prescription of
ref.~\cite{Nicolai:1978vc}.}
\begin{eqnarray}
   S&=&a^4\sum_x\Bigl\{{1\over2}\chi^TCD\chi-{1\over a}X^TCX
   +g(\chi^T+X^T)C\Bigl(\phi+{m\over2g}\Bigr)P_+(\chi+X)
\nonumber\\
   &&\qquad\qquad\qquad\qquad\qquad\qquad\qquad
   +g^*(\chi^T+X^T)C\Bigl(\phi^*+{m^*\over2g^*}\Bigr)P_-(\chi+X)\Bigr\}.
\label{twoxsix}
\end{eqnarray}
When $m=0$, the action is still invariant under the chiral transformation
\begin{eqnarray}
   &&\delta_\alpha\chi=i\alpha\gamma_5(1-{1\over2}aD)\chi
   +i\alpha\gamma_5X,\qquad
   \delta_\alpha X=i\alpha\gamma_5{1\over2}aD\chi,
\nonumber\\
   &&\delta_\alpha\phi=-2i\alpha\phi,\qquad
   \delta_\alpha\phi^*=2i\alpha\phi^*.
\label{twoxseven}
\end{eqnarray}
Eq.~(\ref{twoxsix}) provides a part of our lattice Wess-Zumino model, kinetic
terms of fermions and Yukawa interaction terms.

We next introduce bosonic superpartners of fermion fields, $(\phi,F)$
and~$(\Phi,\mathcal{F})$, and seek an appropriate free action which is
invariant under a certain ``lattice supersymmetry". As the form of this
fermionic transformation, we postulate
\begin{eqnarray}
   &&\delta_\epsilon\chi=-\sqrt{2}P_+(D_1\phi+F)\epsilon
   -\sqrt{2}P_-(D_1\phi^*+F^*)\epsilon,
\nonumber\\
   &&\delta_\epsilon\phi
   =\sqrt{2}\epsilon^TCP_+\chi,\qquad
   \delta_\epsilon\phi^*=\sqrt{2}\epsilon^TCP_-\chi,
\nonumber\\
   &&\delta_\epsilon F=\sqrt{2}\epsilon^TCD_1P_+\chi,\qquad
   \delta_\epsilon F^*=\sqrt{2}\epsilon^TCD_1P_-\chi,
\label{twoxeight}
\end{eqnarray}
and
\begin{eqnarray}
   &&\delta_\epsilon X=-\sqrt{2}P_+(D_1\Phi+\mathcal{F})\epsilon
   -\sqrt{2}P_-(D_1\Phi^*+\mathcal{F}^*)\epsilon,
\nonumber\\
   &&\delta_\epsilon\Phi
   =\sqrt{2}\epsilon^TCP_+X,\qquad
   \delta_\epsilon\Phi^*=\sqrt{2}\epsilon^TCP_-X,
\nonumber\\
   &&\delta_\epsilon\mathcal{F}=\sqrt{2}\epsilon^TCD_1P_+X,\qquad
   \delta_\epsilon\mathcal{F}^*=\sqrt{2}\epsilon^TCD_1P_-X.
\label{twoxnine}
\end{eqnarray}
In this expression, $\epsilon$ is a 4~component Grassmann parameter and we have
used a decomposition of the Dirac operator, $D=D_1+D_2$, where
\begin{equation}
   D_1={1\over2}\gamma_\mu(\partial_\mu^*+\partial_\mu)(A^\dagger A)^{-1/2},
   \qquad
   D_2={1\over a}\Bigl\{1-
   (1+{1\over2}a^2\partial_\mu^*\partial_\mu)(A^\dagger A)^{-1/2}\Bigr\}.
\label{twoxten}
\end{equation}
Note that with respect to spinor space, $D_1$ and~$D_2$ have different
structures. In particular, we have $\{\gamma_5,D_1\}=0$ and~$[\gamma_5,D_2]=0$.
In terms of this decomposition, the Ginsparg-Wilson relation is expressed as
\begin{equation}
   2D_2=a(-D_1^2+D_2^2),
\label{twoxeleven}
\end{equation}
and as a consequence, we have relations
\begin{equation}
   \gamma_5(1-{1\over2}aD)\gamma_5(1-{1\over2}aD)=1-{1\over2}aD_2,\qquad
   \gamma_5(1-{1\over2}aD)\gamma_5D=D_1,
\label{twoxtwelve}
\end{equation}
which will frequently be used below. It is also understood that the $4\times 4$
identity matrix in operators $D_1^2$ and~$D_2$ is omitted when these operators
are acting on bosonic fields. It is then straightforward to see that the
following free action is invariant under~eqs.~(\ref{twoxeight})
and~(\ref{twoxnine}):
\begin{eqnarray}
   S_0&=&a^4\sum_x\Bigl\{{1\over2}\chi^TCD\chi
   +\phi^*D_1^2\phi+F^*F+FD_2\phi+F^*D_2\phi^*
\nonumber\\
   &&\qquad\qquad
   -{1\over a}X^TCX
   -{2\over a}(\mathcal{F}\Phi+\mathcal{F}^*\Phi^*)
\nonumber\\
   &&\qquad\qquad
   +{1\over2}m\tilde\chi^TCP_+\tilde\chi
   +{1\over2}m^*\tilde\chi^TCP_-\tilde\chi
   +m\tilde F\tilde\phi+m^*\tilde F^*\tilde\phi^*
   \Bigr\},
\end{eqnarray}
where we have introduced abbreviations
\begin{equation}
   \tilde\chi=\chi+X,\qquad\tilde\phi=\phi+\Phi,\qquad
   \tilde F=F+\mathcal{F}.
\label{twoxfourteen}
\end{equation}
The combinations~$(\chi,\phi,F)$ and $(X,\Phi,\mathcal{F})$ are regarded as
chiral multiplet in the lattice model. In particular, we refer
$(X,\Phi,\mathcal{F})$ to as the auxiliary chiral multiplet which is
characteristic in the present lattice formulation.

We note that the free action~$S_0$ with $m=0$ possesses three types of $\U(1)$
symmetry~\cite{Aoyama:1998in}. The first is a rather trivial one acting only
on bosonic fields and is defined by the transformation:
\begin{eqnarray}
   &&\delta_\alpha\chi=0,\qquad\delta_\alpha X=0,
\nonumber\\
   &&\delta_\alpha\phi=i\alpha\phi,\qquad
   \delta_\alpha\Phi=i\alpha\Phi,
\nonumber\\
   &&\delta_\alpha F=-i\alpha F,\qquad
   \delta_\alpha\mathcal{F}=-i\alpha\mathcal{F},
\label{twoxfifteen}
\end{eqnarray}
where $\alpha$ is an infinitesimal real parameter. This remains the symmetry
of~$S_0$ even for~$m\neq0$. The second one is nothing but the L\"uscher chiral
symmetry, (\ref{twoxseven}) with $\delta_\alpha\phi=0$,
$\delta_\alpha\Phi=0$, $\delta_\alpha F=0$ and~$\delta_\alpha\mathcal{F}=0$.
Thirdly, somewhat surprisingly, the {\it bosonic\/} sector of $S_0$ with $m=0$
possesses an analogous $\U(1)$ symmetry to eq.~(\ref{twoxseven}):
\begin{eqnarray}
   &&\delta_\alpha\chi=0,\qquad
   \delta_\alpha X=0,
\nonumber\\
   &&\delta_\alpha\phi
   =+i\alpha\{(1-{1\over2}aD_2)\phi-{1\over2}aF^*\}+i\alpha\Phi,\qquad
   \delta_\alpha\Phi=+i\alpha\{{1\over2}aD_2\phi+{1\over2}aF^*\},
\nonumber\\
   &&\delta_\alpha F=+i\alpha\{(1-{1\over2}aD_2)F-{1\over2}aD_1^2\phi^*\}
   +i\alpha\mathcal{F},\qquad
   \delta_\alpha\mathcal{F}=+i\alpha\{{1\over2}aD_2F+{1\over2}aD_1^2\phi^*\},
\nonumber\\
\label{twoxsixteen}
\end{eqnarray}
due to the Ginsparg-Wilson relation. The lattice action~$S_0$ is not invariant
under a uniform rotation of the complex phase of bosonic fields, $\phi$, $F$,
$\Phi$ and~$\mathcal{F}$, due to the presence of terms~$FD_2\phi$
and~$F^*D_2\phi^*$. The above provides a lattice counterpart of this uniform
phase rotation of bosonic fields under which the free action~$S_0$ with~$m=0$
is invariant. Using a linear combination of the above three $\U(1)$ symmetries,
it is possible to define the $\U(1)_R$ symmetry~\cite{Aoyama:1998in} in the
interacting system, as we will see below. It is worthwhile to note that a sum
of transformations (\ref{twoxfifteen}) and (\ref{twoxsixteen}) takes the
following simple form when acting on tilded variables:
\begin{equation}
   \delta_\alpha\tilde\chi=+i\alpha\gamma_5\tilde\chi,\qquad
   \delta_\alpha\tilde\phi=+i\alpha\tilde\phi,\qquad
   \delta_\alpha\tilde F=+i\alpha\tilde F.
\label{twoxseventeen}
\end{equation}

Next we postulate a form of the interaction term as
\begin{equation}
   S_{\rm int.}=a^4\sum_x\Bigl\{
   g\tilde\chi^TC\tilde\phi P_+\tilde\chi+g\tilde F\tilde\phi^2
   +g^*\tilde\chi^TC\tilde\phi^*P_-\tilde\chi
   +g^*\tilde F^*\tilde\phi^{*2}\Bigr\},
\end{equation}
where we have defined interaction terms by taking tilded
variables~(\ref{twoxfourteen}) as unit, because in this way we can relate our
formulation to the Fujikawa-Ishibashi formulation. This way of construction of
interaction terms is also a natural generalization of the Yukawa interaction
in eq.~(\ref{twoxone}). We then find that the full action $S=S_0+S_{\rm int.}$
is ``almost'' invariant under the fermionic transformations~(\ref{twoxeight})
and~(\ref{twoxnine}). In fact, after some algebra using the Fierz
identity,\footnote{We use the identity
$(\chi^TC\chi)\chi=-(\chi^TC\gamma_5\chi)\gamma_5\chi$.} we obtain
\begin{eqnarray}
   \delta_\epsilon S
   &=&-a^4\sum_x\tilde\chi^TC\sqrt{2}\Bigl\{
   gP_+(2\tilde\phi D_1\tilde\phi-D_1\tilde\phi^2)\epsilon
   +g^*P_-(2\tilde\phi^*D_1\tilde\phi^*-D_1\tilde\phi^{*2})\epsilon
   \Bigr\}
\nonumber\\
   &\equiv&-a^4\sum_x\tilde\chi^TC\Delta L\epsilon.
\label{twoxnineteen}
\end{eqnarray}
We emphasize that this breaking could vanish if the Leibniz rule was valid for
the lattice difference operator~$D_1$. In summary, the lattice action for the
Wess-Zumino model
\begin{eqnarray}
   S&=&a^4\sum_x\Bigl\{{1\over2}\chi^TCD\chi
   +\phi^*D_1^2\phi+F^*F+FD_2\phi+F^*D_2\phi^*
\nonumber\\
   &&\qquad\qquad
   -{1\over a}X^TCX
   -{2\over a}(\mathcal{F}\Phi+\mathcal{F}^*\Phi^*)
\nonumber\\
   &&\qquad\qquad
   +{1\over2}m\tilde\chi^TCP_+\tilde\chi
   +{1\over2}m^*\tilde\chi^TCP_-\tilde\chi
   +m\tilde F\tilde\phi+m^*\tilde F^*\tilde\phi^*
\nonumber\\
   &&\qquad\qquad
   +g\tilde\chi^TC\tilde\phi P_+\tilde\chi+g\tilde F\tilde\phi^2
   +g^*\tilde\chi^TC\tilde\phi^*P_-\tilde\chi
   +g^*\tilde F^*\tilde\phi^{*2}
   \Bigr\},
\label{twoxtwenty}
\end{eqnarray}
is invariant under the lattice super transformation~(\ref{twoxeight})
and~(\ref{twoxnine}) up to the breaking term~(\ref{twoxnineteen}).

For the action~(\ref{twoxtwenty}), we can define two types of exact global
``symmetries''. The first is eq.~(\ref{twoxfifteen}) which yields on tilded
variables
\begin{equation}
   \delta_\alpha\tilde\chi=0,\qquad
   \delta_\alpha\tilde\phi=+i\alpha\tilde\phi,\qquad
   \delta_\alpha\tilde F=-i\alpha\tilde F.
\label{twoxtwentyone}
\end{equation}
This is not a symmetry when $g\neq0$, but may be regarded as a ``symmetry'' if
we simultaneously rotate the coupling constant according to
\begin{equation}
   \delta_\alpha g=-i\alpha g.
\label{twoxtwentytwo}
\end{equation}

Another is a lattice counterpart of the $\U(1)_R$ symmetry which is given by a
linear combination of the above three $\U(1)$ transformations:
\begin{eqnarray}
   &&\delta_\alpha\chi=+i\alpha\gamma_5(1-{1\over2}aD)\chi
   +i\alpha\gamma_5X,\qquad
   \delta_\alpha X=+i\alpha\gamma_5{1\over2}aD\chi,
\nonumber\\
   &&\delta_\alpha\phi
   =-3i\alpha\phi+i\alpha\{(1-{1\over2}aD_2)\phi-{1\over2}aF^*\}
   +i\alpha\Phi,
\nonumber\\
   &&\delta_\alpha\Phi=-3i\alpha\Phi
   +i\alpha\{{1\over2}aD_2\phi+{1\over2}aF^*\},
\nonumber\\
   &&\delta_\alpha F=+3i\alpha F
   +i\alpha\{(1-{1\over2}aD_2)F-{1\over2}aD_1^2\phi^*\}
   +i\alpha\mathcal{F},
\nonumber\\
   &&\delta_\alpha\mathcal{F}=+3i\alpha\mathcal{F}
   +i\alpha\{{1\over2}aD_2F+{1\over2}aD_1^2\phi^*\}.
\label{twoxtwentythree}
\end{eqnarray}
On tilded variables, this $\U(1)_R$ transformation takes a simple form
\begin{equation}
   \delta_\alpha\tilde\chi=+i\alpha\gamma_5\tilde\chi,\qquad
   \delta_\alpha\tilde\phi=-2i\alpha\tilde\phi,\qquad
   \delta_\alpha\tilde F=+4i\alpha\tilde F.
\label{twoxtwentyfour}
\end{equation}
The action~$S$ with $m=0$ is invariant under this transformation and this may
also be regarded as a ``symmetry'' even for~$m\neq0$ if we transform the mass
parameter according to
\begin{equation}
   \delta_\alpha m=-2i\alpha m.
\label{twoxtwentyfive}
\end{equation}
Eqs.~(\ref{twoxtwentythree}), (\ref{twoxtwentyfour}) and~(\ref{twoxtwentyfive})
correspond to the $\U(1)_R$ transformation of the continuum Wess-Zumino model.
In fact, if we define the $\U(1)_R$ transformation of the parameter of the
super transformation as
\begin{equation}
   \delta_\alpha\epsilon=-3i\alpha\gamma_5\epsilon,
\end{equation}
then from eqs.~(\ref{twoxeight}) and~(\ref{twoxnine}) it can be confirmed that
$[\delta_\epsilon,\delta_\alpha]=0$ holds on all field variables. The above two
``symmetries'' play an important role when we consider a structure of radiative
corrections in the present model.

In our non-singular local action~(\ref{twoxtwenty}) with interactions, the
chiral $\U(1)_R$ symmetry is realized as an exact symmetry. The no-go theorem
of ref.~\cite{Fujikawa:2002is} on a chiral invariant Yukawa interaction of the
Majorana fermion is evaded in our formulation due to an introduction of the
auxiliary field(s). We further clarify this point in the next subsection.

\subsection{Perturbative equivalence to the Fujikawa-Ishibashi formulation}
In perturbation theory, the above system~$S$ is completely equivalent to a
lattice Wess-Zumino model formulated in
refs.~\cite{Fujikawa:2001ka,Fujikawa:2001ns,Fujikawa:2002ic}, i.e.,
eq.~(\ref{onexone}). A {\it formal\/} way to see this equivalence is to rewrite
the action~$S$ in favor of tilded variables and of the auxiliary multiplet
$(X,\Phi,\mathcal{F})$:
\begin{eqnarray}
   S&=&a^4\sum_x\Bigl\{{1\over2}(\tilde\chi^T-X^T)CD(\tilde\chi-X)
   +(\tilde\phi^*-\Phi^*)D_1^2(\tilde\phi-\Phi)
\nonumber\\
   &&\qquad\qquad
   +(\tilde F^*-\mathcal{F}^*)(\tilde F-\mathcal{F})
   +(\tilde F-\mathcal{F})D_2(\tilde\phi-\Phi)
   +(\tilde F^*-\mathcal{F}^*)D_2(\tilde\phi^*-\Phi^*)
\nonumber\\
   &&\qquad\qquad
   -{1\over a}X^TCX
   -{2\over a}(\mathcal{F}\Phi+\mathcal{F}^*\Phi^*)
\nonumber\\
   &&\qquad\qquad
   +{1\over2}m\tilde\chi^TCP_+\tilde\chi
   +{1\over2}m^*\tilde\chi^TCP_-\tilde\chi
   +m\tilde F\tilde\phi+m^*\tilde F^*\tilde\phi^*
\nonumber\\
   &&\qquad\qquad
   +g\tilde\chi^TC\tilde\phi P_+\tilde\chi+g\tilde F\tilde\phi^2
   +g^*\tilde\chi^TC\tilde\phi^*P_-\tilde\chi
   +g^*\tilde F^*\tilde\phi^{*2}
   \Bigr\}.
\end{eqnarray}
If we perform integrations over the auxiliary chiral multiplet, $X$, $\Phi$
and~$\mathcal{F}$, we have the effective action for tilded variables:
\begin{eqnarray}
   \tilde S&=&
   a^4\sum_x\Bigl\{{1\over2}\tilde\chi^TC
   (1-{1\over2}aD_2)^{-1}D_1\tilde\chi
   -{2\over a}\tilde\phi^*D_2\tilde\phi
   +\tilde F^*(1-{1\over2}aD_2)^{-1}\tilde F
\nonumber\\
   &&\qquad\qquad
   +{1\over2}m\tilde\chi^TCP_+\tilde\chi
   +{1\over2}m^*\tilde\chi^TCP_-\tilde\chi
   +m\tilde F\tilde\phi+m^*\tilde F^*\tilde\phi^*
\nonumber\\
   &&\qquad\qquad
   +g\tilde\chi^TC\tilde\phi P_+\tilde\chi+g\tilde F\tilde\phi^2
   +g^*\tilde\chi^TC\tilde\phi^*P_-\tilde\chi
   +g^*\tilde F^*\tilde\phi^{*2}
   \Bigr\}.
\label{twoxtwentyeight}
\end{eqnarray}
This is, if we identify tilded variables as basic field variables, nothing but
the action~(\ref{onexone}). Associated to the integration over the auxiliary
chiral multiplet, we have
\begin{equation}
   \Pf\Bigl\{C(D-{2\over a})\Bigr\}
   \det\nolimits^{-1}\Bigl\{{2\over a}(D_2-{2\over a})\Bigr\},
\label{twoxtwentynine}
\end{equation}
where the first factor comes from an integration over the 4~component fermionic
spinor~$X$ and the second comes from an integration over complex bosonic
scalars $\Phi$ and~$\mathcal{F}$; the operator in the latter factor therefore
does not contain $4\times4$ identity matrix in spinor space. In a formal sense,
these two factors are cancelled out, because the relation
\begin{equation}
   \gamma_5(D-{2\over a})\gamma_5(D-{2\over a})
   =-{2\over a}(D_2-{2\over a})
\label{twoxthirty}
\end{equation}
holds (recall eq.~(\ref{twoxtwelve})) and thus we have
\begin{equation}
   \det\nolimits^2\Bigl\{D-{2\over a}\Bigr\}
   =\det\nolimits^4\Bigl\{{-2\over a}(D_2-{2\over a})\Bigr\},
\end{equation}
by noting the fact that the right hand side of eq.~(\ref{twoxthirty}) contains
the $4\times 4$ identity matrix. Therefore, we see
\begin{equation}
   \Pf\Bigl\{C(D-{2\over a})\Bigr\}
   =\det\nolimits^{1/2}C\det\nolimits^{1/2}\Bigl\{D-{2\over a}\Bigr\}
   =\det\nolimits^{1/2}C
   \det\nolimits\Bigl\{{-2\over a}(D_2-{2\over a})\Bigr\}.
\end{equation}
This cancels the contribution from the bosonic fields. The system~$S$ is thus
equivalent to~$\tilde S$ after integrating out the auxiliary chiral multiplet,
$X$, $\Phi$ and~$\mathcal{F}$.

This argument for the equivalence between $S$ and~$\tilde S$, however, is valid
in a formal sense, because the operators $D-{2\over a}$ and thus
$D_2-{2\over a}$ which appear in various places in the above expressions always
have zero modes when the lattice volume is
infinite~\cite{Fujikawa:2001ns,Fujikawa:2002is}. The kinetic operators of~$X$
and of~$(\Phi,\mathcal{F})$ have zero eigenmodes {\it when tilded variables are
kept fixed}. Thus the integration over the former gives zero and the latter
gives infinity. On the other hand, kinetic operators in the effective
action~$\tilde S$ contain the factor~$(1-{1\over2}aD_2)^{-1}$ which is a
singular operator.

To see what is really happening here, it is instructive to consider the case
of~$m=g=0$. In this case, integrations over Grassmann variables yield
\begin{equation}
   \int\prod_x\rmd\chi(x)\rmd X(x)\,
   e^{-a^4\sum_x\{{1\over2}\chi^TCD\chi-{1\over a}X^TCX\}}
   =\Pf\{CD\}\Pf\Bigl\{-{2\over a}C\Bigr\},
\label{twoxthirtythree}
\end{equation}
which is {\it not\/} singular in any sense. On the other hand, if we perform
the integration over~$X$ first {\it while keeping $\tilde\chi=\chi+X$}, we have
instead
\begin{equation}
   \Pf\Bigl\{C(D-{2\over a})\Bigr\}
   \int\prod_x\rmd\tilde\chi(x)\,
   e^{-a^4\sum_x\{{1\over2}\tilde\chi^TC(1-{1\over2}aD_2)^{-1}D_1\tilde\chi\}},
\end{equation}
which is of a structure of~$0\times\infty$, although nothing is wrong with the
whole integral~(\ref{twoxthirtythree}). The above argument simply shows that we
are observing a non-singular object in an unnecessarily singular way. A similar
argument is applied to the full action~(\ref{twoxtwenty}). Integrations over
field variables do not produce any singularities. Only if we observe the
integrations in a wrong way, seemingly singular natures as in
eqs.~(\ref{twoxtwentyeight}) and~(\ref{twoxtwentynine}) emerge. The full action
$S$~(\ref{twoxtwenty}) and the action $\tilde S$~(\ref{twoxtwentyeight}) are
thus {\it inequivalent\/} by a singular quantity.\footnote{In this way, the
no-go theorem of ref.~\cite{Fujikawa:2002is} for a chiral invariant Yukawa
interaction of the Majorana fermion is evaded by introducing auxiliary fields.}

Nevertheless, we can infer that our formulation based on~$S$ and that based
on~$\tilde S$ are equivalent in {\it perturbation theory}. The point is that
the free propagators among {\it tilded\/} variables
\begin{eqnarray}
   &&\langle\tilde\chi(x)\tilde\chi^T(y)\rangle C
   ={-D_1+(1-{1\over2}aD_2)(m^*P_++mP_-)
   \over{2\over a}D_2+(1-{1\over2}aD_2)m^*m}
   a^{-4}\delta_{x,y},
\nonumber\\
   &&\langle\tilde\phi(x)\tilde\phi^*(y)\rangle
   =\langle\tilde\phi^*(x)\tilde\phi(y)\rangle
   ={-1
   \over{2\over a}D_2+(1-{1\over2}aD_2)m^*m}a^{-4}\delta_{x,y},
\nonumber\\
   &&\langle\tilde\phi(x)\tilde F(y)\rangle
   =\langle\tilde F(x)\tilde\phi(y)\rangle
   ={(1-{1\over2}aD_2)m^*
   \over{2\over a}D_2+(1-{1\over2}aD_2)m^*m}a^{-4}\delta_{x,y},
\nonumber\\
   &&\langle\tilde\phi^*(x)\tilde F^*(y)\rangle
   =\langle\tilde F^*(x)\tilde\phi^*(y)\rangle
   ={(1-{1\over2}aD_2)m
   \over{2\over a}D_2+(1-{1\over2}aD_2)m^*m}a^{-4}\delta_{x,y},
\nonumber\\
   &&\langle\tilde F(x)\tilde F^*(y)\rangle
   =\langle\tilde F^*(x)\tilde F(y)\rangle
   ={(1-{1\over2}aD_2){2\over a}D_2
   \over{2\over a}D_2+(1-{1\over2}aD_2)m^*m}a^{-4}\delta_{x,y},
\label{twoxthirtyfive}
\end{eqnarray}
which are directly obtained by using~$S$, are identical to those obtained by
using~$\tilde S$ formally, i.e., by neglecting a singular nature of kinetic
terms. Moreover, interaction vertices of~$S$ are identical to those
of~$\tilde S$ (in fact we have constructed $S$ so that this is the case).
Therefore, perturbative calculations based on~$S$ and that based on~$\tilde S$
give rise to completely identical answers for correlation functions which have
tilded variables in external lines. In a sense, our non-singular local lattice
action~$S$ provides a natural justification for a prescription of
refs.~\cite{Fujikawa:2001ka,Fujikawa:2002ic} which utilizes the above form of
propagators and interaction vertices of~$\tilde S$. Of course, we think our
formulation which includes the auxiliary chiral supermultiplet is superior at
least formally, because it is manifestly free from singularities and it may
have a meaning even as a non-perturbative formulation.

\section{Supersymmetric Ward-Takahashi identity and its breaking}
\subsection{Derivation of a lattice Ward-Takahashi identity}
We consider a structure of radiative corrections with the present lattice
formulation of the Wess-Zumino model. As noted in the preceding section, in
perturbation theory, our formulation is equivalent to the formulation of
refs.~\cite{Fujikawa:2001ka,Fujikawa:2001ns,Fujikawa:2002ic}. One-loop
radiative corrections in the latter formulation, in view of a realization of
supersymmetry, had been extensively studied in ref.~\cite{Fujikawa:2001ka}.
Here we study this issue in the continuum limit by using a Ward-Takahashi
identity.

For a systematic study, it is quite helpful to introduce the one-particle
irreducible (1PI) effective action~$\Gammait$. Following the standard
procedure, we introduce external sources for each elementary fields
\begin{equation}
   S_{\rm source}=a^4\sum_x\{J_\chi\chi+J_\phi\phi+J_{\phi^*}\phi^*
   +J_FF+J_{F^*}F^*
   +J_XX+J_\Phi\Phi+J_{\Phi^*}\Phi^*
   +J_{\mathcal{F}}\mathcal{F}+J_{\mathcal{F}^*}\mathcal{F}^*\}.
\end{equation}
We also introduce an external source~$K$ for a symmetry breaking of the action,
$\delta_\epsilon S$, and a source~$L$ for a symmetry variation of
$\delta_\epsilon S$, $\delta_{\epsilon'}\delta_\epsilon S$, and so on.
Including these latter kind of external sources $K$, $L$\dots\ only, we define
the {\it total action}
\begin{equation}
   S_{\rm tot.}
   =S-a^4\sum\{K_\alpha(\tilde\chi^TC\Delta L)_\alpha
   +L_{\alpha,\beta}\delta_\alpha(\tilde\chi^TC\Delta L)_\beta
   +M_{\alpha\beta,\gamma}\delta_\alpha\delta_\beta
   (\tilde\chi^TC\Delta L)_\gamma
   +\cdots\},
\label{threextwo}
\end{equation}
where $\delta_\alpha$ stands for the symmetry variation with the transformation
spinor parameter is removed:
\begin{equation}
   \delta_\epsilon=\epsilon_\alpha\delta_\alpha.
\end{equation}
The generating functional~$W$ of connected Green's functions is then defined by
the functional integral
\begin{eqnarray}
   e^{-W}&=&\int\prod_x\rmd\chi(x)\rmd\phi(x)\rmd\phi^*(x)
   \rmd F(x)\rmd F^*(x)
   \rmd X(x)\rmd\Phi(x)\rmd\Phi^*(x)
   \rmd\mathcal{F}(x)\rmd\mathcal{F}^*(x)
\nonumber\\
   &&\qquad\qquad
   \times e^{-S_{\rm tot.}-S_{\rm source}}.
\end{eqnarray}
We then apply the Legendre transformation to~$W$ and change independent
variables from external sources for elementary fields $(J_\chi,\cdots)$ to the
corresponding expectation values of elementary fields
$(\langle\chi\rangle,\cdots)$. In what follows, we denote expectation values by
their original name as $\langle\chi\rangle\to\chi$ and so on for notational
simplicity. We do {\it not\/} apply the Legendre transformation with respect to
the sources $(K,L,M,\ldots)$. In this way, we have the 1PI effective action
\begin{equation}
   \Gammait=\Gammait[\chi,\phi,\phi^*,F,F^*,X,\Phi,\Phi^*,
   \mathcal{F},\mathcal{F}^*;K,L,M,\ldots],
\end{equation}
which is a generating functional of 1PI Green's functions which include
additional vertices coming from the second term of eq.~(\ref{threextwo}).

Now, the action~$S$ is not invariant under the lattice super
transformation~(\ref{twoxeight}) and~(\ref{twoxnine}), but it leaves the
breaking~(\ref{twoxnineteen}). From this fact, we have the identity
\begin{eqnarray}
   &&\Bigl\langle
   -a^4\sum_x\{J_\chi\delta_\epsilon\chi
   +J_\phi\delta_\epsilon\phi
   +J_{\phi^*}\delta_\epsilon\phi^*
   +J_F\delta_\epsilon F+J_{F^*}\delta_\epsilon F^*
\nonumber\\
   &&\qquad\qquad
   +J_X\delta_\epsilon X
   +J_\Phi\delta_\epsilon\Phi
   +J_{\Phi^*}\delta_\epsilon\Phi^*
   +J_{\mathcal{F}}\delta_\epsilon\mathcal{F}
   +J_{\mathcal{F}^*}\delta_\epsilon\mathcal{F}^*\}
\nonumber\\
   &&\quad+a^4\sum_x\tilde\chi^TC\Delta L\epsilon
\nonumber\\
   &&\quad+a^4\sum\{
   K_\alpha\epsilon_\beta\delta_\beta(\tilde\chi^TC\Delta L)_\alpha
   +L_{\alpha,\beta}\epsilon_\gamma\delta_\gamma
   \delta_\alpha(\tilde\chi^TC\Delta L)_\beta
   +\cdots\}\Bigr\rangle=0.
\label{threexsix}
\end{eqnarray}
This identity, in terms of the 1PI effective action~$\Gammait$, is expressed as
\begin{eqnarray}
   &&-\sum_x\Gammait{\overleftarrow\partial\over\partial\chi}
   \{\sqrt{2}P_+(D_1\phi+F)\epsilon+\sqrt{2}P_-(D_1\phi^*+F^*)\epsilon\}
\nonumber\\
   &&+\sum_x\Gammait{\overleftarrow\partial\over\partial\phi}\sqrt{2}
   \epsilon^TCP_+\chi
   +\sum_x\Gammait{\overleftarrow\partial\over\partial\phi^*}\sqrt{2}
   \epsilon^TCP_-\chi
\nonumber\\
   &&+\sum_x\Gammait{\overleftarrow\partial\over\partial F}\sqrt{2}
   \epsilon^TCD_1P_+\chi
   +\sum_x\Gammait{\overleftarrow\partial\over\partial F^*}\sqrt{2}
   \epsilon^TCD_1P_-\chi
\nonumber\\
   &&-\sum_x\Gammait{\overleftarrow\partial\over\partial X}
   \{\sqrt{2}P_+(D_1\Phi+\mathcal{F})\epsilon
   +\sqrt{2}P_-(D_1\Phi^*+\mathcal{F}^*)\epsilon\}
\nonumber\\
   &&+\sum_x\Gammait{\overleftarrow\partial\over\partial\Phi}\sqrt{2}
   \epsilon^TCP_+X
   +\sum_x\Gammait{\overleftarrow\partial\over\partial\Phi^*}\sqrt{2}
   \epsilon^TCP_-X
\nonumber\\
   &&+\sum_x\Gammait{\overleftarrow\partial\over\partial\mathcal{F}}\sqrt{2}
   \epsilon^TCD_1P_+X
   +\sum_x\Gammait{\overleftarrow\partial\over\partial\mathcal{F}^*}\sqrt{2}
   \epsilon^TCD_1P_-X
\nonumber\\
   &&+\sum_x{\partial\over\partial K_\alpha}\Gammait
   \epsilon_\alpha
\nonumber\\
   &&+\sum_x\Bigl\{K_\alpha\epsilon_\beta
   {\partial\over\partial L_{\beta,\alpha}}\Gammait
   +L_{\alpha,\beta}\epsilon_\gamma
   {\partial\over\partial M_{\gamma\alpha,\beta}}\Gammait
   +\cdots\Bigr\}=0,
\label{threexseven}
\end{eqnarray}
that is a linear equation of~$\Gammait$. This is referred to as the
{\it lattice\/} Ward-Takahashi identity. If the last two lines were not present
in this expression, the above equation simply states that the effective action
is invariant under the lattice analogue of super transformation,
eqs.~(\ref{twoxeight}) and~(\ref{twoxnine}). Thus, contributions of these
lines, especially contributions from the term,
\begin{equation}
   \sum_x{\partial\over\partial K_\alpha}\Gammait\epsilon_\alpha,
\label{threexeight}
\end{equation}
that is the breaking of the supersymmetric Ward-Takahashi identity, will play a
central role in our analysis below. Explicitly, this term is given by 1PI
diagrams with insertions of the operator\footnote{Generally, $\Gammait$
contains 1PI diagrams with multiple insertions of this operator. We will be
interested in, however, terms of~$\Gammait$ that are linear in the external
source~$K$ and consider 1PI diagrams with a single insertion of this operator
below.}
\begin{eqnarray}
   &&-a^4\sum_x\tilde\chi^TC\Delta L\epsilon
\nonumber\\
   &&=-a^4\sum_x\tilde\chi^TC\sqrt{2}\Bigl\{
   gP_+(2\tilde\phi D_1\tilde\phi-D_1\tilde\phi^2)\epsilon
   +g^*P_-(2\tilde\phi^*D_1\tilde\phi^*-D_1\tilde\phi^{*2})\epsilon
   \Bigr\}.
\label{threexnine}
\end{eqnarray}
The $a\to0$ limit of these 1PI diagrams will be expressed by a {\it local\/}
polynomial of field variables, because in this limit, the effect of our
particular choice of regularization (the lattice artifact) should affect only
local terms in the effective action~$\Gammait$. Moreover, the
operator~(\ref{threexnine}) vanishes in the classical continuum limit (because
the Leibniz rule holds in this limit) and it has no continuum analogue. Only
when combined with ultraviolet divergences,
$\langle\tilde\chi^TC\Delta L\epsilon\rangle_{\rm 1PI}$ can acquire non-zero
value. In these aspects, computation
of~$\langle\tilde\chi^TC\Delta L\epsilon\rangle_{\rm 1PI}$ is similar to that
of quantum anomalies, although this breaking of supersymmetry is not a genuine
anomaly in a conventional sense.\footnote{It will be removed by local counter
terms (supersymmetry is thought to be anomaly-free at least in perturbation
theory) and also the structure
of~$\langle\tilde\chi^TC\Delta L\epsilon\rangle_{\rm 1PI}$ is not universal,
i.e., it will quite depend on a lattice formulation one adopts.}

We expand $\Gammait$ according to a number of internal loops of 1PI diagrams:
\begin{equation}
   \Gammait=\Gammait_0+\Gammait_1+\Gammait_2+\cdots.
\end{equation}
The loop counting parameter in the present model is~$g^*g$. The tree-level
effective action~$\Gammait_0$ is nothing but the total
action~(\ref{threextwo}),
\begin{equation}
   \Gammait_0=S_{\rm tot.}
\end{equation}
In fact, it is easy to see that $S_{\rm tot.}$ satisfies the lattice
Ward-Takahashi identity. In this tree level approximation, the breaking term
vanishes in the $a\to0$ limit,
\begin{equation}
   \lim_{a\to0}\tilde\chi^TC\Delta L\epsilon=0,
\end{equation}
because the Leibniz rule holds in this limit. Thus the last two lines of the
identity~(\ref{threexseven}) vanish in the $a\to0$ limit and the supersymmetry
is restored in this classical continuum limit.

In loop diagrams, all external lines are tilded variables. The Ward-Takahashi
identity for the effective action~$\Gammait_n$ $(n\geq1)$ can thus be written
as
\begin{eqnarray}
   &&-\sum_x\Gammait_n{\overleftarrow\partial\over\partial\tilde\chi}
   \{\sqrt{2}P_+(D_1\tilde\phi+\tilde F)\epsilon
   +\sqrt{2}P_-(D_1\tilde\phi^*+\tilde F^*)\epsilon\}
\nonumber\\
   &&+\sum_x\Gammait_n{\overleftarrow\partial\over\partial\tilde\phi}\sqrt{2}
   \epsilon^TCP_+\tilde\chi
   +\sum_x\Gammait_n{\overleftarrow\partial\over\partial\tilde\phi^*}\sqrt{2}
   \epsilon^TCP_-\tilde\chi
\nonumber\\
   &&+\sum_x\Gammait_n{\overleftarrow\partial\over\partial\tilde F}\sqrt{2}
   \epsilon^TCD_1P_+\tilde\chi
   +\sum_x\Gammait_n{\overleftarrow\partial\over\partial\tilde F^*}\sqrt{2}
   \epsilon^TCD_1P_-\tilde\chi
\nonumber\\
   &&+\sum_x{\partial\over\partial K_\alpha}\Gammait_n
   \epsilon_\alpha
\nonumber\\
   &&+\sum_x\Bigl\{K_\alpha\epsilon_\beta
   {\partial\over\partial L_{\beta,\alpha}}\Gammait_n
   +L_{\alpha,\beta}\epsilon_\gamma
   {\partial\over\partial M_{\gamma\alpha,\beta}}\Gammait_n
   +\cdots\Bigr\}=0.
\label{threexthirteen}
\end{eqnarray}

We also recall that our system~$S$ possesses two global $\U(1)$ ``symmetries'':
$\U(1)$ symmetry, eqs.~(\ref{twoxtwentyone}) and~(\ref{twoxtwentytwo}),
and $\U(1)_R$ symmetry, eqs.~(\ref{twoxtwentyfour}) and~(\ref{twoxtwentyfive}).
In terms of the 1PI effective action $\Gammait_n$ $(n\geq1)$, these invariance
can be expressed as
\begin{equation}
   \sum_x\Bigl\{
   \Gammait_n{\overleftarrow\partial\over\partial\tilde\phi}\tilde\phi
   -\Gammait_n{\overleftarrow\partial\over\partial\tilde\phi^*}\tilde\phi^*
   -\Gammait_n{\overleftarrow\partial\over\partial\tilde F}\tilde F
   +\Gammait_n{\overleftarrow\partial\over\partial\tilde F^*}\tilde F^*\Bigr\}
   -\Gammait_n{\overleftarrow\partial\over\partial g}g
   +\Gammait_n{\overleftarrow\partial\over\partial g^*}g^*=0,
\end{equation}
and
\begin{eqnarray}
   &&\sum_x\Bigl\{
   \Gammait_n{\overleftarrow\partial\over\partial\tilde\chi}\gamma_5\tilde\chi
   -2\Gammait_n{\overleftarrow\partial\over\partial\tilde\phi}\tilde\phi
   +2\Gammait_n{\overleftarrow\partial\over\partial\tilde\phi^*}\tilde\phi^*
   +4\Gammait_n{\overleftarrow\partial\over\partial\tilde F}\tilde F
   -4\Gammait_n{\overleftarrow\partial\over\partial\tilde F^*}\tilde F^*\Bigr\}
\nonumber\\
   &&\qquad\qquad\qquad\qquad\qquad\qquad\qquad\qquad
   -2\Gammait_n{\overleftarrow\partial\over\partial m}m
   +2\Gammait_n{\overleftarrow\partial\over\partial m^*}m^*=0,
\end{eqnarray}
where we have set $K=L=\cdots=0$. These identities are referred to as
Ward-Takahashi identities associated to $\U(1)$ symmetries.

\subsection{Supersymmetric Ward-Takahashi identity: Improvement and
renormalization}
Supersymmetry is not exact in the present lattice formulation of the
Wess-Zumino model. To achieve a supersymmetric continuum limit, we have to
apply appropriate improvement and renormalization to the lattice action. In
particular, in the continuum limit, a renormalized effective action must obey
the {\it supersymmetric\/} Ward-Takahashi identity that is defined by
\begin{eqnarray}
   &&-\int\rmd^4x\,
   \Gammait_n{\overleftarrow\delta\over\delta\tilde\chi}
   \{\sqrt{2}P_+(\gamma_\mu\partial_\mu\tilde\phi+\tilde F)\epsilon
   +\sqrt{2}P_-(\gamma_\mu\partial_\mu\tilde\phi^*+\tilde F^*)\epsilon\}
\nonumber\\
   &&+\int\rmd^4x\,
   \Gammait_n{\overleftarrow\partial\over\partial\tilde\phi}\sqrt{2}
   \epsilon^TCP_+\tilde\chi
   +\int\rmd^4x\,
   \Gammait_n{\overleftarrow\partial\over\partial\tilde\phi^*}\sqrt{2}
   \epsilon^TCP_-\tilde\chi
\nonumber\\
   &&+\int\rmd^4x\,
   \Gammait_n{\overleftarrow\partial\over\partial\tilde F}\sqrt{2}
   \epsilon^TC\gamma_\mu\partial_\mu P_+\tilde\chi
   +\int\rmd^4x\,
   \Gammait_n{\overleftarrow\partial\over\partial\tilde F^*}\sqrt{2}
   \epsilon^TC\gamma_\mu\partial_\mu P_-\tilde\chi=0,
\label{threexsixteen}
\end{eqnarray}
and\footnote{This requirement may seem somewhat ad hoc. However, without this
kind of additional condition, divergences arise from sub-diagrams in a 1PI
diagram which contains source terms $K$, $L$, $M$, \dots\ cannot be cancelled
by an addition of local terms to $\Gammait_{n+1}$. Note that this requirement
is satisfied in the tree level.}
\begin{equation}
   \int\rmd^4x\,{\delta\over\delta K_\alpha}\Gammait_n
   =\int\rmd^4x\,{\delta\over\delta L_{\beta,\alpha}}\Gammait_n
   =\int\rmd^4x\,{\delta\over\delta M_{\gamma\alpha,\beta}}\Gammait_n
   =\cdots=0.
\label{threexseventeen}
\end{equation}
Also the $\U(1)$ symmetries must be preserved
\begin{equation}
   \int\rmd^4x\,\Bigl\{
   \Gammait_n{\overleftarrow\partial\over\partial\tilde\phi}\tilde\phi
   -\Gammait_n{\overleftarrow\partial\over\partial\tilde\phi^*}\tilde\phi^*
   -\Gammait_n{\overleftarrow\partial\over\partial\tilde F}\tilde F
   +\Gammait_n{\overleftarrow\partial\over\partial\tilde F^*}\tilde F^*\Bigr\}
   -\Gammait_n{\overleftarrow\partial\over\partial g}g
   +\Gammait_n{\overleftarrow\partial\over\partial g^*}g^*=0,
\label{threexeighteen}
\end{equation}
and
\begin{eqnarray}
   &&\int\rmd^4x\,\Bigl\{
   \Gammait_n{\overleftarrow\partial\over\partial\tilde\chi}\gamma_5\tilde\chi
   -2\Gammait_n{\overleftarrow\partial\over\partial\tilde\phi}\tilde\phi
   +2\Gammait_n{\overleftarrow\partial\over\partial\tilde\phi^*}\tilde\phi^*
   +4\Gammait_n{\overleftarrow\partial\over\partial\tilde F}\tilde F
   -4\Gammait_n{\overleftarrow\partial\over\partial\tilde F^*}\tilde F^*\Bigr\}
\nonumber\\
   &&\qquad\qquad\qquad\qquad\qquad\qquad\qquad\qquad
   -2\Gammait_n{\overleftarrow\partial\over\partial m}m
   +2\Gammait_n{\overleftarrow\partial\over\partial m^*}m^*=0.
\label{threexnineteen}
\end{eqnarray}
In these expressions, all fields, external sources, mass parameters and
coupling constants are regarded as {\it renormalized\/} quantities. To achieve
this supersymmetric finite continuum theory, we take following steps. (1)~We
compute 1PI $n$th loop diagrams $\Gammait_n$ by using the total action
$S_{\rm tot.}$. (2)~We add appropriate local counter terms to the total
action~$S_{\rm tot.}$ so that the supersymmetric Ward-Takahashi
identities~(\ref{threexsixteen}) and~(\ref{threexseventeen}) hold in the
$a\to0$ limit. At this stage, all fields and parameters in
eqs.~(\ref{threexsixteen}) and~(\ref{threexseventeen}) are understood as bare
quantities. This improvement step removes supersymmetry breaking due to lattice
artifacts in our formulation. (3)~Then we further add appropriate local counter
terms to the total action~$S_{\rm tot.}$ so that $\Gammait_n$ is finite in the
$a\to0$ limit. This step corresponds to the standard supersymmetric
renormalization and we assume its validity, i.e., we assume a renormalizability
of this lattice model. Explicitly, we modify the total action as 
\begin{eqnarray}
   S_{\rm tot.}&\to&S_{\rm tot.}
   +a^4\sum_xZ_n\Bigl\{
   {1\over2}\chi^TCD\chi
   +\phi^*D_1^2\phi+F^*F
\nonumber\\
   &&\qquad\qquad\qquad
   +FD_2\phi+F^*D_2\phi^*
   -{1\over a}X^TCX
   -{2\over a}(\mathcal{F}\Phi+\mathcal{F}^*\Phi^*)\Bigr\},
\label{threextwenty}
\end{eqnarray}
where $Z_n$ is a common wave function renormalization factor. (4)~All these
steps must be consistent with global $\U(1)$ symmetries. We then repeat the
above steps for 1PI diagrams of one loop higher, $\Gammait_{n+1}$, by using
$S_{\rm tot.}$ so determined.

The tree-level effective action~$\Gammait_0$ is given by the total action in
eq.~(\ref{threextwo}) and of course it does not need any renormalization and
improvement. To see the situation in the one-loop level, we evaluate in the
next subsection the supersymmetry breaking term in the lattice Ward-Takahashi
identity~(\ref{threexthirteen}) to this order.

\subsection{One-loop evaluation of the breaking term}
We evaluate the supersymmetry breaking term in eq.~(\ref{threexthirteen}) in
the one-loop order. We set $K=L=\cdots=0$. It is given by one-loop 1PI diagrams
with a single insertion of the operator~(\ref{threexnine}). A computation of
the continuum limit of these one-loop 1PI diagrams is not difficult, if one
invokes a powerful Reisz's theorem~\cite{Reisz:1987da,Luscher:1988sd} on
lattice Feynman integrals.

Most singular one-loop diagrams which possibly contribute to
\begin{equation}
   \lim_{a\to0}
   \sum_x{\partial\over\partial K_\alpha}\Gammait_1\epsilon_\alpha
   \Bigr|_{K=L=\cdots=0}
   =\lim_{a\to0}\Bigl\{
   -a^4\sum_x\langle\tilde\chi^TC\Delta L\rangle_{\rm 1PI}^{\rm 1loop}
   \epsilon\Bigr|_{K=L=\cdots=0}\Bigr\},
\label{threextwentyone}
\end{equation}
are given by figures~1--4 and their complex conjugate.
\EPSFIGURE[t]{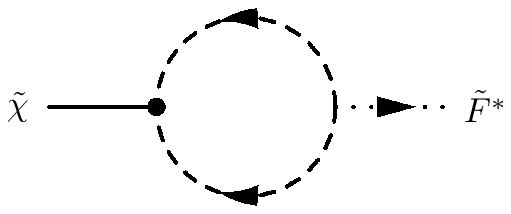,width=.6\textwidth}{A diagram which contributes to
eq.~(\ref{threextwentyone}). The blob indicates the supersymmetry breaking
operator~(\ref{threexnine}). The broken line is the propagator of~$\tilde\phi$
and the arrow indicates a flow of the chirality.}
\EPSFIGURE[t]{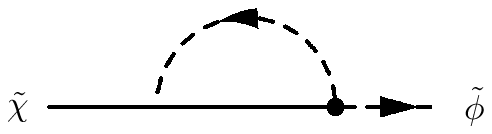,width=.6\textwidth}{A diagram which possibly contributes
to eq.~(\ref{threextwentyone}). The bold line is the propagator
of~$\tilde\chi$. The $a\to0$ limit of this diagram turns out to be vanishing.}
\EPSFIGURE[t]{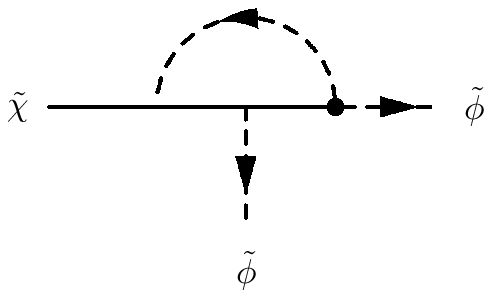,width=.6\textwidth}{A diagram which possibly contributes
to eq.~(\ref{threextwentyone}). The $a\to0$ limit of this diagram turns out to
be vanishing due to the $\U(1)_R$ symmetry of the $m=0$ case.}
\EPSFIGURE[t]{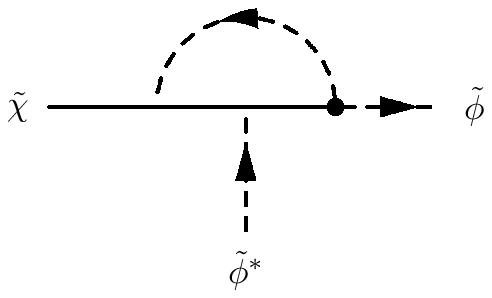,width=.6\textwidth}{A diagram which possibly contributes
to eq.~(\ref{threextwentyone}). The $a\to0$ limit of this diagram turns out to
be vanishing due to the $\U(1)_R$ symmetry of the $m=0$ case.}

By using propagators in eq.~(\ref{twoxthirtyfive}), the contribution from
figure~1 to eq.~(\ref{threextwentyone}) is given by
\begin{equation}
   +2\sqrt{2}g^*ga^4\sum_xa^4\sum_y\tilde\chi^T(x)CP_+\tilde F^*(y)
   \int_{\mathcal{B}}{\rmd^4q\over(2\pi)^4}\,e^{iq(x-y)}
   \int_{\mathcal{B}}{\rmd^4k\over(2\pi)^4}\,
   \mathcal{I}_1(k,q;m,a)\epsilon,
\end{equation}
where $\mathcal{B}$ denotes the Brillouin zone
\begin{equation}
   \mathcal{B}=\left\{p\in R^4\mid|p_\mu|\leq\pi/a\right\},
\end{equation}
and the integrand is given by
\begin{eqnarray}
   &&\mathcal{I}_1(k,q;m,a)
\nonumber\\
   &&=
   {1\over{2\over a}\widetilde D_2(k)
   +\left(1-{1\over2}a\widetilde D_2(k)\right)m^*m}
   {\widetilde D_1(k+q)
   -\widetilde D_1(k)-\widetilde D_1(q)\over{2\over a}\widetilde D_2(k+q)
   +\left(1-{1\over2}a\widetilde D_2(k+q)\right)m^*m}.
\end{eqnarray}
In this expression, $\widetilde D_1$ and $\widetilde D_2$ denote the momentum
representation of difference operators in eq.~(\ref{twoxten})
\begin{equation}
   D_1e^{ipx}=\widetilde D_1(p)e^{ipx},\qquad
   D_2e^{ipx}=\widetilde D_2(p)e^{ipx},
\end{equation}
and the explicit forms are given by
\begin{eqnarray}
   &&\widetilde D_1(p)
   =i\sum_\mu\gamma_\mu\ring p_\mu
   \Bigl\{1
   +{1\over2}a^4\sum_{\nu<\rho}\hat p_\nu^2\hat p_\rho^2\Bigr\}^{-1/2},
\nonumber\\
   &&\widetilde D_2(p)
   ={1\over a}\Bigl\{1-
   \Bigl(1-{1\over2}a^2\sum_\mu\hat p_\mu^2\Bigr)
   \Bigl\{1
   +{1\over2}a^4\sum_{\nu<\rho}\hat p_\nu^2\hat p_\rho^2\Bigr\}^{-1/2}
   \Bigr\},
\end{eqnarray}
with abbreviations
\begin{equation}
   \ring p_\mu={1\over a}\sin(ap_\mu),\qquad
   \hat p_\mu={2\over a}\sin\left({ap_\mu\over2}\right).
\end{equation}
We note that both $\ring p_\mu$ and~$\hat p_\mu$ reduce to the momentum in the
continuum theory in the $a\to0$ limit, $\lim_{a\to0}\ring p_\mu=%
\lim_{a\to0}\hat p_\mu=p_\mu$.

Now, a crucial idea of the Reisz power counting theorem is to consider the
$\lambda\to\infty$ limit of the integrand, after replacing the internal loop
momenta $k_i$ by $\lambda k_i$ and the lattice spacing~$a$ by $a/\lambda$. From
the above explicit form of the integrand, we find
\begin{equation}
   \mathcal{I}_1(\lambda k,q;m,a/\lambda)=O(1/\lambda^4),
\end{equation}
in the $\lambda\to\infty$ limit. This implies that the degree of divergence of
the above loop integral (in a sense of the Reisz power counting theorem) is~$0$
and the $a\to0$ limit of the loop integral may exhibit a logarithmic divergence
of the form~$\log a$.

To reduce the degree of divergence, we thus take the first term in the Taylor
expansion of the integrand with respect to the external momentum~$q$ and
consider a subtraction of the form
\begin{equation}
   \mathcal{I}_1(k,q;m,a)-\mathcal{I}_1(k,0;m,a).
\end{equation}
However, since $\mathcal{I}_1(k,0;m,a)=0$, this subtraction does not improve
the convergence behavior at all.

We are thus lead to consider a subtraction to the next order term in the Taylor
expansion
\begin{equation}
   \mathcal{I}_1(k,q;m,a)-\mathcal{I}_1(k,0;m,a)
   -q_\mu\partial_\mu^q\mathcal{I}_1(k,0;m,a).
\end{equation}
Then we find that this combination behaves as $O(1/\lambda^6)$ in the
$\lambda\to\infty$ limit defined above. The Reisz power counting theorem then
states that the $a\to0$ of the loop integral of this combination is convergent
and moreover the limit is given by\footnote{In ref.~\cite{Bartels:1982ue}, this
fact is referred to as an empirical ``rule''.}
\begin{eqnarray}
   &&\lim_{a\to0}\int_{\mathcal{B}}{\rmd^4k\over(2\pi)^4}\,\left\{
   \mathcal{I}_1(k,q;m,a)-\mathcal{I}_1(k,0;m,a)
   -q_\mu\partial_\mu^q\mathcal{I}_1(k,0;m,a)\right\}
\nonumber\\
   &&=\int_{R^4}{\rmd^4k\over(2\pi)^4}\,\lim_{a\to0}\left\{
   \mathcal{I}_1(k,q;m,a)-\mathcal{I}_1(k,0;m,a)
   -q_\mu\partial_\mu^q\mathcal{I}_1(k,0;m,a)\right\}
\nonumber\\
   &&=0.
\label{threexthirtyone}
\end{eqnarray}
The last equality follows from a property of the
operator~$\Delta L$~(\ref{threexnine}) that it vanishes in the classical
continuum limit.

In this way, we obtain
\begin{eqnarray}
   &&\lim_{a\to0}\int_{\mathcal{B}}{\rmd^4k\over(2\pi)^4}\,
   \mathcal{I}_1(k,q;m,a)
\nonumber\\
   &&=\lim_{a\to0}\int_{\mathcal{B}}{\rmd^4k\over(2\pi)^4}\,\left\{
   \mathcal{I}_1(k,0;m,a)
   +q_\mu\partial_\mu^q\mathcal{I}_1(k,0;m,a)\right\}
\nonumber\\
   &&\qquad+\lim_{a\to0}\int_{\mathcal{B}}{\rmd^4k\over(2\pi)^4}\,\left\{
   \mathcal{I}_1(k,q;m,a)-\mathcal{I}_1(k,0;m,a)
   -q_\mu\partial_\mu^q\mathcal{I}_1(k,0;m,a)\right\}
\nonumber\\
   &&=\int_{-\pi}^\pi{\rmd^4k\over(2\pi)^4}\,\lim_{a\to0}
   {1\over a^4}q_\mu\partial_\mu^q\mathcal{I}_1(k/a,0;m,a)
\nonumber\\
   &&=-{r_1\over2}i\gamma_\mu q_\mu,
\label{threexthirtytwo}
\end{eqnarray}
where the coefficient~$r_1$ is given by an integral
\begin{eqnarray}
   r_1&=&{-1\over32\pi^4}\int_{-\pi}^\pi\rmd^4k\,
   \Bigl\{\Bigl(1-{1\over2}\hat k_0^2\Bigr)B^{-1/2}
   -{1\over2}\ring k_0^2\sum_{\mu\neq0}\hat k_\mu^2\,B^{-3/2}-1\Bigr\}
\nonumber\\
   &&\qquad\qquad\qquad\qquad\qquad\qquad\qquad
   \times\Bigl\{1-\Bigl(1-{1\over2}\sum_\mu\hat k_\mu^2\Bigr)
   B^{-1/2}\Bigr\}^{-2},
\end{eqnarray}
with abbreviations
\begin{equation}
   \ring k_\mu=\sin k_\mu,\qquad\hat k_\mu=2\sin({k_\mu\over2}),\qquad
   B=1+{1\over2}\sum_{\mu<\nu}\hat k_\mu^2\hat k_\nu^2.
\end{equation}
From a numerical integration, we have
\begin{equation}
   r_1=+0.1518.
\end{equation}
Eq.~(\ref{threexthirtytwo}) then implies that figure~1 gives rise to
\begin{eqnarray}
   &&\lim_{a\to0}\Bigl\{
   -a^4\sum_x\langle\tilde\chi^TC\Delta L\rangle_{\rm 1PI}^{\rm 1loop}
   \epsilon\Bigr|_{K=L=\cdots=0}\Bigr\}
\nonumber\\
   &&=-r_1g^*g\int\rmd^4x\,\Bigl\{
   \tilde F^*\sqrt{2}\epsilon^TC\gamma_\mu\partial_\mu P_+\tilde\chi
   +\tilde F\sqrt{2}\epsilon^TC\gamma_\mu\partial_\mu P_-\tilde\chi\Bigr\}.
\label{threexthirtyfive}
\end{eqnarray}

Next, we consider the contribution of figure~2 which is given by
\begin{equation}
   +4\sqrt{2}g^*ga^4\sum_xa^4\sum_y\tilde\chi^T(x)CP_-\tilde\phi(y)
   \int_{\mathcal{B}}{\rmd^4q\over(2\pi)^4}\,e^{iq(x-y)}
   \int_{\mathcal{B}}{\rmd^4k\over(2\pi)^4}\,
   \mathcal{I}_2(k,q;m,a)\epsilon,
\end{equation}
where
\begin{eqnarray}
   &&\mathcal{I}_2(k,q;m,a)
\nonumber\\
   &&=
   {\widetilde D_1(k)\over{2\over a}\widetilde D_2(k)
   +\left(1-{1\over2}a\widetilde D_2(k)\right)m^*m}
   {\widetilde D_1(k+q)
   -\widetilde D_1(k)-\widetilde D_1(q)\over{2\over a}\widetilde D_2(k+q)
   +\left(1-{1\over2}a\widetilde D_2(k+q)\right)m^*m}.
\end{eqnarray}
We find that $\mathcal{I}_2(\lambda k,q;m,a/\lambda)=O(1/\lambda^3)$ in the
$\lambda\to\infty$ limit and the degree of divergence of the above integral
is~$1$. A twice subtraction
\begin{equation}
   \mathcal{I}_2(k,q;m,a)-\mathcal{I}_2(k,0;m,a)
   -q_\mu\partial_\mu^q\mathcal{I}_2(k,0;m,a),
\label{threexthirtynine}
\end{equation}
then makes the degree of divergence~$-1$ and the $a\to0$ limit of the loop
integral convergent. Again, as before, the $a\to0$ limit of the
combination~(\ref{threexthirtynine}) vanishes. However, this time,
$\mathcal{I}_2(k,0;m,a)=0$ and the remaining lattice integral vanishes too:
\begin{eqnarray}
   \int_{\mathcal{B}}{\rmd^4k\over(2\pi)^4}\,
   q_\mu\partial_\mu^q\mathcal{I}_2(k,0;m,a)=0.
\end{eqnarray}
Thus figure~2 has no contribution in the $a\to0$ limit.

The contribution of figure~3 is given by
\begin{eqnarray}
   &&-8\sqrt{2}m^*g^*g^2a^4\sum_xa^4\sum_ya^4\sum_z
   \tilde\chi^T(x)CP_-\tilde\phi(y)\tilde\phi(z)
   \int_{\mathcal{B}}{\rmd^4q_1\over(2\pi)^4}\,e^{iq_1(x-z)}
   \int_{\mathcal{B}}{\rmd^4q_2\over(2\pi)^4}\,e^{iq_2(x-y)}
\nonumber\\
   &&\qquad\qquad\qquad\qquad\qquad\qquad\qquad\qquad\qquad
   \times\int_{\mathcal{B}}{\rmd^4k\over(2\pi)^4}\,
   \mathcal{I}_3(k,q_1,q_2;m,a)\epsilon,
\end{eqnarray}
where
\begin{eqnarray}
   &&\mathcal{I}_3(k,q_1,q_2;m,a)
\nonumber\\
   &&={\widetilde D_1(k-q_2)
   \over{2\over a}\widetilde D_2(k-q_2)
   +\left(1-{1\over2}a\widetilde D_2(k-q_2)\right)m^*m}
   {1-{a\over2}\widetilde D_2(k)
   \over{2\over a}\widetilde D_2(k)
   +\left(1-{1\over2}a\widetilde D_2(k)\right)m^*m}
\nonumber\\
   &&\qquad\qquad\qquad\qquad\qquad
   \times{\widetilde D_1(k+q_1)-\widetilde D_1(k)-\widetilde D_1(q_1)
   \over{2\over a}\widetilde D_2(k+q_1)
   +\left(1-{1\over2}a\widetilde D_2(k+q_1)\right)m^*m}.
\end{eqnarray}
This behaves as $O(1/\lambda^5)$ in the $\lambda\to\infty$ limit defined above
(the degree of divergence is~$-1$) and the $a\to0$ limit of the loop integral
converges without any subtraction. Since
$\lim_{a\to0}\mathcal{I}_3(k,q;m,a)=0$, however, the contribution of figure~3
vanishes in the $a\to0$ limit. An underlying reason for this good convergence
behavior of figure~3 is the chiral $\U(1)_R$ symmetry of~$S$ with~$m=0$ under
eq.~(\ref{twoxtwentythree}). Due to this symmetry, this diagram vanishes for
$m=0$ even with $a\neq0$. In fact, if this diagram had a contribution to the
breaking term when~$m=0$, through the Ward-Takahashi
identity~(\ref{threexthirteen}), there must be terms of the form
$\tilde\phi^*\tilde\phi^2$ or $\tilde F^*\tilde\phi^2$ in the one-loop
effective action~$\Gammait_1$: Both are however forbidden by the chiral
$\U(1)_R$ symmetry~(\ref{twoxtwentyfour}). Similarly we find that figure~4 has
no contribution in the $a\to0$ limit.

\EPSFIGURE[t]{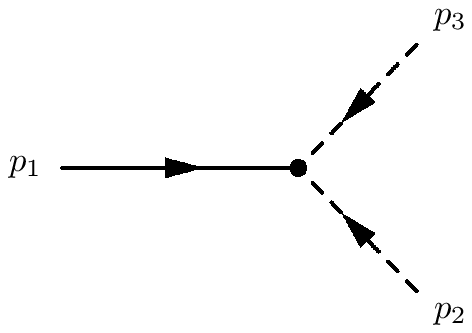,width=.4\textwidth}{The supersymmetry breaking
vertex~(\ref{threexnine}) in the momentum space.}
As is also clear from above expressions, the vertex for the breaking term
$\tilde\chi^TC\Delta L\epsilon$ in eq.~(\ref{threexnine}) in the momentum
space~(see figure~5) is proportional to a
combination~$\widetilde D_1(p_1)+\widetilde D_1(p_2)+\widetilde D_1(p_3)$. From
the fact that in one-loop diagrams one of three legs of the vertex is always
an external line and from the momentum conservation at the vertex, we find that
this vertex, when inserted in a one-loop diagram, gives rise to $O(\lambda^0)$
factor in the $\lambda\to\infty$ limit defined above. This implies that this
vertex effectively behaves as a non-derivative coupling in the power counting
argument for one-loop diagrams. Then it is easy to confirm that the degree of
divergence of a loop integral in all possible one-loop diagrams other than
those in figures~1--4 is negative. By repeating a similar argument as above, we
then infer that the $a\to0$ limit of those contributions is zero.

In summary, only the contribution of figure~1 survives in the $a\to0$ limit
and we have
\begin{eqnarray}
   &&\lim_{a\to0}
   \sum_x{\partial\over\partial K_\alpha}\Gammait_1\epsilon_\alpha
   \Bigr|_{K=L=\cdots=0}
\nonumber\\
   &&=-r_1g^*g\int\rmd^4x\,\Bigl\{
   \tilde F^*\sqrt{2}\epsilon^TC\gamma_\mu\partial_\mu P_+\tilde\chi
   +\tilde F\sqrt{2}\epsilon^TC\gamma_\mu\partial_\mu P_-\tilde\chi\Bigr\}.
\label{threexfortythree}
\end{eqnarray}
As a general argument shows, the supersymmetry breaking in the $a\to0$ limit is
a local polynomial of fields.

\subsection{One-loop level improvement and the renormalization}
From the above one-loop calculation~(\ref{threexfortythree}), we can extract
following information. First, combining it with the $a\to0$ limit of the
lattice Ward-Takahashi identity~(\ref{threexthirteen}), we find that the
coefficient of the term~$\int\rmd^4x\,\tilde F^*\tilde F$ in~$\Gammait_1$ is
different from the supersymmetric value. Namely, the coefficient of this term
is not consistent with the supersymmetric Ward-Takahashi
identity~(\ref{threexsixteen}). Next, eq.~(\ref{threexfortythree}) shows that
$\Gammait_1$ contains a finite term of the form~$\int\rmd^4x\,\{
\tilde F^*\sqrt{2}K^TC\gamma_\mu\partial_\mu P_+\tilde\chi+{\rm c.c.}\}$ which
is not consistent with eq.~(\ref{threexseventeen}). These two are {\it only\/}
places in $\Gammait_1$ in which the breaking of supersymmetry appears in the
continuum limit (to the order~$O(K,L^0,\ldots)$). According to the general
strategy of section~3.2, we thus modify the total action~$S_{\rm tot.}$ to
\begin{eqnarray}
   &&S_{\rm tot.}-a^4\sum_x r_1g^*g\Bigl\{
   \tilde F^*\tilde F
   +\tilde F^*\sqrt{2}K^TCD_1 P_+\tilde\chi
   +\tilde F\sqrt{2}K^TCD_1 P_-\tilde\chi\Bigr\}
\nonumber\\
   &&\qquad\qquad\qquad\qquad\qquad\qquad\qquad\qquad\qquad\qquad
   +O(K^2,L,\ldots).
\label{threexfortyfour}
\end{eqnarray}
to restore the supersymmetric Ward-Takahashi identity~(\ref{threexsixteen}).
The added terms contribute to~$\Gammait_1$.

In ref.~\cite{Fujikawa:2001ka}, one-loop 1PI two point functions are computed
and it was found that radiative corrections to kinetic terms of $\tilde\chi$,
$\tilde\phi$ and~$\tilde F$ are in general different by finite amount, although
logarithmically diverging parts are of an identical magnitude. Our observation
above is consistent with this result and is slightly stronger: We observed that
only the wave function renormalization of the auxiliary field~$\tilde F$
differs from other two in the continuum limit. The improvement above adjusts
this discrepancy in wave function renormalization factors.

After the above improvement, the effective action~$\Gammait_1$ becomes
supersymmetric. The standard statements concerning the supersymmetric
Wess-Zumino model are then applied to the one-loop effective
action~$\Gammait_1$. For example, wave function renormalization factors for
$\tilde\chi$, $\tilde\phi$ and~$\tilde F$ are common as indicated in
eq.~(\ref{threextwenty}). Also, a local term of the form
\begin{equation}
   \int\rmd^4x\,V(\tilde\phi^*,\tilde\phi),
\end{equation}
where $V$ is an arbitrary local polynomial of $\tilde\phi$ and~$\tilde\phi^*$
without any derivatives, does not appear in the one-loop effective
action~$\Gammait_1$, simply because such a combination is not supersymmetric,
i.e, it is not a solution to the supersymmetric Ward-Takahashi
identity~(\ref{threexsixteen}).\footnote{An exception in this argument is
$V={\rm const.}$ that is nothing but the cosmological term. One cannot exclude
the cosmological term from the supersymmetric Ward-Takahashi identity alone.
In the present model, the cosmological term vanishes in the one-loop level
(see appendix~A), as expected from an {\it exact\/} supersymmetry of the free
action~$S_0$.} This conclusion is again consistent with a one-loop analysis of
ref.~\cite{Fujikawa:2001ka} that no terms consisting only of $\phi$
and~$\phi^*$ are generated by one-loop radiative corrections.\footnote{The
result of ref.~\cite{Fujikawa:2001ka} is somewhat stronger: Up to the quartic
order in $\tilde\phi$ or $\tilde\phi^*$, it was observed that such terms are
not generated even for $a\neq0$.}

By a similar reasoning, we can also show the non-renormalization
theorem~\cite{Iliopoulos:1974zv}--\cite{Grisaru:1979wc} of the form in
ref.~\cite{Grisaru:1979wc}. The non-renormalization theorem states that the
$F$~term of the structure
\begin{equation}
   \int\rmd^4x\,\Bigl\{
   {1\over2}\tilde\chi^TC W''(\tilde\phi)P_+\tilde\chi
   +\tilde FW'(\tilde\phi)\Bigr\},
\label{threexfortysix}
\end{equation}
where $W(\tilde\phi)$ is the superpotential, is not generated by radiative
corrections, although this {\it is\/} a supersymmetric combination. We first
note that, from a structure of one-loop diagrams, dependences of such a term
on the coupling constant must be of the form
\begin{equation}
   \int\rmd^4x\,\{g\tilde FW'(g\tilde\phi)+\cdots\}.
\label{threexfortyseven}
\end{equation}
The complex conjugate~$g^*$ cannot appear here from a structure of interaction
vertices. Next we recall that our lattice action~$S$ possesses an exact $\U(1)$
symmetry under eqs.~(\ref{twoxtwentyone}) and~(\ref{twoxtwentytwo}). The above
allowed structure~(\ref{threexfortyseven}) is, however, inconsistent with this
$\U(1)$ symmetry.\footnote{Our argument here is somewhat similar to that of
ref.~\cite{Seiberg:1993vc}. The conclusion here, however, is just a reflection
of a simple fact that there is {\it no\/} 1PI one-loop diagram made out from
only $\tilde F$ and~$\tilde\phi$ external lines.} Thus we conclude that the
$F$~term~(\ref{threexfortysix}) as the whole cannot be generated by one-loop
radiative corrections. This conclusion is again consistent with the analysis of
ref.~\cite{Fujikawa:2001ka}; there it was observed that terms such
as~$\tilde\chi^TC\tilde\chi$ and~$\tilde\chi^TC\tilde\phi P_+\tilde\chi$ are
not generated in the one-loop order.

Finally, by further adding local counter terms~(\ref{threextwenty}), the
one-loop effective action~$\Gammait_1$ is made finite, i.e., a supersymmetric
renormalized theory is defined. Obviously the theory so defined preserves
the $\U(1)$ symmetries, i.e., eqs.~(\ref{threexeighteen})
and~(\ref{threexnineteen}) hold, because all stages of the above procedure
(and the lattice regularization itself) do not affect these symmetries.

\subsection{Higher loops}
In the one-loop order, we have observed that if we add local terms of the
form~(\ref{threexfortyfour}) to the total action, especially by an adjustment
of the term~$\tilde F^*\tilde F$, supersymmetry is restored in the continuum
limit. In this subsection, we consider if this simple situation persists to
higher orders of perturbation theory. We will find that, instead of a single
combination of local terms~(\ref{threexfortyfour}), there are at most
9~combinations that we have to take into account for an improvement of higher
loop effective action~$\Gammait_n$. The improvement in higher loops is thus
much more involved. We note, however, this number of combinations which require
adjustments is much less than that in the formulation based on the Wilson
fermion~\cite{Bartels:1982ue} due to the exact $\U(1)_R$
symmetry~(\ref{twoxtwentyfour}) and~(\ref{twoxtwentyfive}) in the present
formulation.

Suppose that the procedure in section~3.2 of the renormalization and the
improvement (which may require an addition of 9~combinations of local terms
to~$S_{\rm tot.}$) work for~$\Gammait_1$, \dots, $\Gammait_{n-1}$. Now take a
1PI $n$~loop diagram~$\gamma_n$ which contains a single insertion of the
operator~(\ref{threexnine}). From the above assumption, all 1PI
{\it sub-diagrams\/} are already made finite by the renormalization of
$\Gammait_1$, \dots, $\Gammait_{n-1}$. By applying a standard power counting
argument to the present case, the superficial degree of divergence of such a
diagram~$\gamma_n$ is given by
\begin{equation}
   \omega(\gamma_n)={9\over 2}-{3\over2}E_\chi-E_\phi-2E_F-I_{\phi F},
\label{threexfortyeight}
\end{equation}
where $E_\chi$, $E_\phi$, $E_F$ denote a number of external lines of
$\tilde\chi$, $\tilde\phi$ or $\tilde\phi^*$, $\tilde F$ or $\tilde F^*$,
respectively and $I_{\phi F}$ denotes a number of $\tilde\phi\tilde F$
and~$\tilde\phi^*\tilde F^*$~type internal lines. To derive this formula, one
has to note that the propagator $\langle\tilde F\tilde F^*\rangle$ behaves as
$O(\lambda^0)$ (no suppression factor) in the Reisz power counting rule.

Also, in deriving the above formula, we have noted a fact that an insertion of
the vertex~(\ref{threexnine}) in higher loop diagrams effectively behaves
as~$O(\lambda)$ in the Reisz power counting rule, because in higher loop
diagrams all momenta in figure~5 can simultaneously become large. In one-loop
diagrams, on the other hand, the vertex~figure~5 behaves as~$O(\lambda^0)$
because one of three legs must always be an external line and momentum from
the external line is kept fixed in the $\lambda\to\infty$ limit. (For one-loop
diagrams, $9/2$ in the formula~(\ref{threexfortyeight}) is changed to~$7/2$;
see below.) This difference in a behavior of the supersymmetry breaking term in
the $\lambda\to\infty$ limit in one-loop and higher-loop diagrams is crucial
and due to this difference much more combinations have to be included in local
counter terms to obtain a supersymmetric continuum limit.

Now, since we have assumed that all 1PI sub-diagrams of~$\gamma_n$ are made
finite by a renormalization, the above superficial degree of divergence will be
an overall degree of divergence. Then if $\omega(\gamma_n)<0$, the Reisz
theorem states that the $a\to0$ limit of the diagram~$\gamma_n$ is given by
$R^4$ integrations of the $a\to0$ limit of the integrand, as we have seen in
eq.~(\ref{threexthirtyone}). However, due to the vertex~(\ref{threexnine}), the
$a\to0$ limit of the integrand always vanishes. Hence diagrams which can
contribute to the supersymmetry breaking in the continuum limit must possess
$\omega(\gamma_n)\geq0$. Noting for $\gamma_n$, $E_\chi=1$, 3, \dots, we can
see that there are seven combinations for~$\omega(\gamma_n)\geq0$, i.e.,
$(E_\chi,E_\phi,E_F)=(1,0,0)$, $(1,1,0)$, $(1,2,0)$, $(1,3,0)$, $(1,0,1)$,
$(1,1,1)$ and~$(3,0,0)$. The total mass dimension of $\gamma_n$ is $9/2$.

A structure of~$\gamma_n$ moreover must be consistent with exact global
symmetries in the present formulation; the $\U(1)_R$ symmetry,
eqs.~(\ref{twoxtwentyfour}) and~(\ref{twoxtwentyfive}), and the $\U(1)$
symmetry, eqs.~(\ref{twoxtwentyone}) and~(\ref{twoxtwentytwo}). The $\U(1)_R$
and the $\U(1)$ charges of $\gamma_n$ are $(-3,+1)$ or $(+3,-1)$.

Finally, we have to check a consistency of~$\gamma_n$ with the lattice
Ward-Takahashi identity~(\ref{threexthirteen}). By examining various possible
local terms in~$\Gammait_n$ which also must be consistent with the global
symmetries, we finally find that a most general form of
\begin{equation}
   \lim_{a\to0}
   \sum_x{\partial\over\partial K_\alpha}\Gammait_n\epsilon_\alpha
   \Bigr|_{K=L=\cdots=0},
\end{equation}
is given by\footnote{To list up possible combinations, it is easier to work out
the massless case~$m=0$ first and then restore possible dependences on~$m$ by
substituting $\tilde\phi\to m/g$ and~$\tilde\phi^*\to m^*/g^*$ in arbitrary
ways.}
\begin{eqnarray}
   &&-(g^*g)^n\int d^4x\,\Bigl\{r_n
   \tilde F^*\sqrt{2}\epsilon^TC\gamma_\mu\partial_\mu P_+\tilde\chi
\nonumber\\
   &&\qquad
   +\Bigl(s_n\partial_\mu^2\tilde\phi^*
   +{1\over a^2}t_n\tilde\phi^*
   +2u_n\tilde\phi^{*2}\tilde\phi
\nonumber\\
   &&\qquad\qquad
   +{1\over a^2}v_n{m^*\over g^*}
   +w_n\Bigl[2{m^*\over g^*}\tilde\phi^*\tilde\phi
   +{m\over g}\tilde\phi^{*2}\Bigr]
   +2x_n{m^{*2}\over g^{*2}}\tilde\phi
   \Bigr)\sqrt{2}\epsilon^TCP_+\tilde\chi
\nonumber\\
   &&\qquad\qquad\qquad
   +y_n(2g\tilde F\tilde\phi\sqrt{2}\epsilon^TCP_+\tilde\chi
   +g\tilde\phi^2\sqrt{2}\epsilon^TC\gamma_\mu\partial_\mu P_+\tilde\chi)
\nonumber\\
   &&\qquad\qquad\qquad\qquad
   +z_n(m\tilde F\sqrt{2}\epsilon^TCP_+\tilde\chi
   +m\tilde\phi\sqrt{2}\epsilon^TC\gamma_\mu\partial_\mu P_+\tilde\chi)
   \Bigr\}
\label{threexfifty}
\end{eqnarray}
plus its complex conjugate (the projection operator is replaced
by~$P_+\to P_-$). The real coefficients $r_n$, \dots, $z_n$ are given by
dimensionless polynomials of $\log(a^2m^*m)$ and~$a^2m^*m$.

Through the $a\to0$ limit of the lattice Ward-Takahashi identity, the above
form of the breaking term implies that the supersymmetry breaking terms in the
effective action take the form:
\begin{eqnarray}
   &&+(g^*g)^n\int d^4x\,\Bigl\{r_n
   \tilde F^*\tilde F
\nonumber\\
   &&\qquad
   +s_n\tilde\phi^*\partial_\mu^2\tilde\phi
   +{1\over a^2}t_n\tilde\phi^*\tilde\phi
   +u_n\tilde\phi^{*2}\tilde\phi^2
\nonumber\\
   &&\qquad\qquad
   +{1\over a^2}v_n\Bigl({m^*\over g^*}\tilde\phi+{m\over g}\tilde\phi^*\Bigr)
   +w_n\Bigl({m^*\over g^*}\tilde\phi^*\tilde\phi^2
   +{m\over g}\tilde\phi^{*2}\tilde\phi\Bigr)
   +x_n\Bigl({m^{*2}\over g^{*2}}\tilde\phi^2
   +{m^2\over g^2}\tilde\phi^{*2}\Bigr)
\nonumber\\
   &&\qquad\qquad\qquad
   +y_n(g\tilde F\tilde\phi^2+g^*\tilde F^*\tilde\phi^{*2})
   +z_n(m\tilde F\tilde\phi+m^*\tilde F^*\tilde\phi^*)
   \Bigr\}
\nonumber\\
   &&+(g^*g)^n\int d^4x\,\Bigl\{r_n
   \tilde F^*\sqrt{2}K^TC\gamma_\mu\partial_\mu P_+\tilde\chi
\nonumber\\
   &&\qquad
   +\Bigl(s_n\partial_\mu^2\tilde\phi^*
   +{1\over a^2}t_n\tilde\phi^*
   +2u_n\tilde\phi^{*2}\tilde\phi
\nonumber\\
   &&\qquad\qquad
   +{1\over a^2}v_n{m^*\over g^*}
   +w_n\Bigl[2{m^*\over g^*}\tilde\phi^*\tilde\phi
   +{m\over g}\tilde\phi^{*2}\Bigr]
   +2x_n{m^{*2}\over g^{*2}}\tilde\phi
   \Bigr)\sqrt{2}K^TCP_+\tilde\chi
\nonumber\\
   &&\qquad\qquad\qquad
   +y_n(2g\tilde F\tilde\phi\sqrt{2}K^TCP_+\tilde\chi
   +g\tilde\phi^2\sqrt{2}K^TC\gamma_\mu\partial_\mu P_+\tilde\chi)
\nonumber\\
   &&\qquad\qquad\qquad\qquad
   +z_n(m\tilde F\sqrt{2}K^TCP_+\tilde\chi
   +m\tilde\phi\sqrt{2}K^TC\gamma_\mu\partial_\mu P_+\tilde\chi)+{\rm c.c.}
   \Bigr\}
\nonumber\\
   &&\qquad\qquad\qquad\qquad\qquad\qquad\qquad\qquad\qquad
   +O(K^2,L,\ldots).
\label{threexfiftyone}
\end{eqnarray}
It is easy to verify that these are, modulo supersymmetric combinations, most
general local terms whose mass dimension $\leq4$ that are consistent with the
$\U(1)_R$ and~$\U(1)$ global symmetries (the $\U(1)$ charges of the external
source~$P_\pm K$ is~$(\mp3,\pm1)$). This is an expected result from an
experience in continuum theory but is not entirely trivial, because the term
which breaks supersymmetry due to a violation of the Leibniz
rule~(\ref{threexnine}) is peculiar to lattice theory and cannot be treated in
a framework of continuum theory.

To remedy the above breaking of supersymmetry, we subtract
eq.~(\ref{threexfiftyone}) from $S_{\rm tot.}$ after transcribing it as local
terms in lattice theory by substitutions
\begin{equation}
   \int\rmd^4 x\to a^4\sum_x,\qquad
   \partial_\mu^2\to D_1^2,\qquad
   \gamma_\mu\partial_\mu\to D_1.
\end{equation}
This is the improvement step; we have to add 9~combinations of local terms
to~$S_{\rm tot.}$ for $\Gammait_n$ to have a supersymmetric continuum limit.
Then a further supersymmetric renormalization~(\ref{threextwenty}) will make
$\Gammait_n$ finite. Although this procedure may be applied in principle, the
number of required local terms for the improvement is too many for
any practical application of the present model.

We can understand why the situation in the one-loop level was so simple by
considering the case in which the first term in eq.~(\ref{threexfortyeight})
is~$7/2$ instated of~$9/2$. Repeating a similar analysis as above, as a
possible form of
$\lim_{a\to0}\sum_x{\partial\over\partial K_\alpha}\Gammait_1\epsilon_\alpha
|_{K=L=\cdots=0}$, we obtain
\begin{equation}
   -g^*g\int d^4x\,\Bigl\{
   r_1\tilde F^*\sqrt{2}\epsilon^TC\gamma_\mu\partial_\mu P_+\tilde\chi
   +s_1\partial_\mu^2\tilde\phi^*
   \sqrt{2}\epsilon^TCP_+\tilde\chi\Bigr\}
\end{equation}
plus the complex conjugate. This has a much simpler structure than
eq.~(\ref{threexfifty}). We in fact found the term with~$r_1$ in the explicit
one-loop calculation~(\ref{threexfortythree}). We also observe that a general
argument does not prohibit a non-zero~$s_1$; thus $s_1=0$ in the explicit
one-loop calculation~(\ref{threexfortythree}) is accidental.

\subsection{Exact non-linear fermionic symmetry}
Finally, we comment on an exact non-linear fermionic symmetry in this system
which corresponds to the symmetry studied in ref.~\cite{Bonini:2004pm}. We
note that the lattice Ward-Takahashi identity~(\ref{threexsix}) (after setting
$K=L=M=\cdots=0$) may be written as
\begin{eqnarray}
   &&\Bigl\langle
   -a^4\sum_x\{J_\chi\delta_\epsilon\chi
   +J_\phi\delta_\epsilon\phi
   +J_{\phi^*}\delta_\epsilon\phi^*
   +J_F\delta_\epsilon F
   +J_{F^*}\delta_\epsilon F^*
\nonumber\\
   &&\qquad\qquad
   +J_X\delta_\epsilon X
   +J_\Phi\delta_\epsilon\Phi
   +J_{\Phi^*}\delta_\epsilon\Phi^*
   +J_{\mathcal{F}}\delta_\epsilon\mathcal{F}
   +J_{\mathcal{F}^*}\delta_\epsilon\mathcal{F}^*\}
\nonumber\\
   &&\quad+\sum_x\Bigl\{S{\overleftarrow\partial\over\partial\chi}
   R\epsilon
   +S{\overleftarrow\partial\over\partial X}
   \mathcal{R}\epsilon\Bigr\}
   \Bigr\rangle=0,
\label{threexfiftyfour}
\end{eqnarray}
where
\begin{eqnarray}
   &&R=\Bigl[D+mP_++m^*P_-
   +2(g\tilde\phi P_++g^*\tilde\phi^*P_-)
   \Bigl\{1-{1\over2}a(D+mP_++m^*P_-)\Bigr\}
   \Bigr]^{-1}\Delta L
\nonumber\\
   &&\mathcal{R}=-{1\over2}a(D+mP_++m^*P_-)R.
\end{eqnarray}
However, noting the Schwinger-Dyson equation
\begin{equation}
   \Bigl\langle
   \sum_xS{\overleftarrow\partial\over\partial\varphi}\delta\varphi
   +a^4\sum_xJ_\varphi\delta\varphi\Bigr\rangle=0,
\end{equation}
where $\varphi$ and $\delta\varphi$ represent a generic field and its
{\it arbitrary\/} variation, the identity~(\ref{threexfiftyfour}) can be cast
into the form
\begin{eqnarray}
   &&\Bigl\langle
   -a^4\sum_x\{J_\chi\delta'_\epsilon\chi
   +J_\phi\delta'_\epsilon\phi
   +J_{\phi^*}\delta'_\epsilon\phi^*
   +J_F\delta'_\epsilon F+J_{F^*}\delta'_\epsilon F^*
\nonumber\\
   &&\qquad\qquad
   +J_X\delta'_\epsilon X
   +J_\Phi\delta'_\epsilon\Phi
   +J_{\Phi^*}\delta'_\epsilon\Phi^*
   +J_{\mathcal{F}}\delta'_\epsilon\mathcal{F}
   +J_{\mathcal{F}^*}\delta'_\epsilon\mathcal{F}^*\}
   \Bigr\rangle=0,
\label{threexfiftyseven}
\end{eqnarray}
where $\delta'_\epsilon$ is a transformation modified by amount of $R$
and~$\mathcal{R}$:
\begin{eqnarray}
   &&\delta'_\epsilon\chi=-\sqrt{2}P_+(D_1\phi+F)\epsilon
   -\sqrt{2}P_-(D_1\phi^*+F^*)\epsilon+R\epsilon,
\nonumber\\
   &&\delta'_\epsilon X=-\sqrt{2}P_+(D_1\Phi+\mathcal{F})\epsilon
   -\sqrt{2}P_-(D_1\Phi^*+\mathcal{F}^*)\epsilon+\mathcal{R}\epsilon,
\nonumber\\
   &&\delta'_\epsilon=\delta_\epsilon\qquad\hbox{on other fields}.
\end{eqnarray}
Obviously, the above form of the identity~(\ref{threexfiftyseven}) can be
regarded as a Ward-Takahashi identity associated to an {\it exact\/}
symmetry~$\delta'_\epsilon$ of the action~$S$. The
transformation~$\delta'_\epsilon$ when acting on tilded variables (i.e., in the
context of the Fujikawa-Ishibashi formulation) is nothing but the exact
non-linear transformation studied in ref.~\cite{Bonini:2004pm}. However, as we
have demonstrated, Ward-Takahashi identities in both pictures, one is based
on~$\delta_\epsilon$ (eq.~(\ref{threexfiftyfour})) and another is based on
$\delta'_\epsilon$ (eq.~(\ref{threexfiftyseven})), have identical
contents.\footnote{Actually, this is the starting point of the argument of
ref.~\cite{Golterman:1988ta}.} The presence of this exact symmetry itself does
not imply the lattice formulation is ``better'' in any sense. Despite this
exact symmetry, an adjustment of parameters is needed to obtain a
supersymmetric continuum limit, as we have discussed in this paper.

\section{Conclusion}
In this paper, we formulated a lattice model for the $N=1$ supersymmetric
Wess-Zumino model in 4~dimensions. The $\U(1)_R$ symmetry is manifest even on
a lattice with a use of Ginsparg-Wilson fermions. Although our formulation is
perturbatively equivalent to the Fujikawa-Ishibashi formulation, we could
avoid a singular nature of the latter formulation by introducing an auxiliary
chiral supermultiplet on a lattice. We also analyzed radiative breaking of the
supersymmetric Ward-Takahashi identity. The situation in the one-loop order is
rather simple while the improvement through higher orders will be much more
involved due to a peculiarity of the supersymmetry breaking term. In
particular, in higher orders, we cannot avoid an adjustment of mass parameters
of scalar fields which are quadratically diverging. In this aspect, the
situation is not quite better than for the formulation based on the Wilson
fermion. Clearly, a much clever idea is needed to achieve a lattice formulation
of the $N=1$ Wess-Zumino model which avoids this too much adjustment.

\acknowledgments
H.S. would like to thank Ke Wu for kind hospitality at Institute of Theoretical
Physics, Academia Sinica, where an integral part of this work was carried out.
This work is supported in part by Grant-in-Aid for Scientific Research,
\#13135203 and \#15540250 (H.S.).

\paragraph{Note added in proofs.}
In an illustrative work~\cite{Giedt:2004vb}, Giedt, Koniuk, Poppitz and Yavin
studied a fine-tuning problem in a naively discretized supersymmetric quantum
mechanics by utilizing the Reisz theorem. We thank Joel Giedt for calling their
work to our attention.

\appendix

\section{Cosmological term at the one-loop level}
The cosmological term (the vacuum energy) at the one-loop level is given by a
logarithm of the partition function with $g=0$. For a correct counting of
degrees of freedom, it is helpful to introduce real bosonic variables by
\begin{eqnarray}
   &&\phi\to(A+iB)/\sqrt{2},\qquad F\to(F-iG)/\sqrt{2},
\nonumber\\
   &&\Phi\to(\alpha+i\beta)/\sqrt{2},\qquad
   \mathcal{F}\to(\mathcal{F}-i\mathcal{G})/\sqrt{2}.
\end{eqnarray}
Then the free part of action~$S_0$ is represented as
\begin{eqnarray}
   S_0&=&a^4\sum_x
   (A\,\,B\,\,F\,\,G\,\,\alpha\,\,\beta\,\,\mathcal{F}\,\,\mathcal{G})
   M_{\rm B}
   \pmatrix{A\cr B\cr F\cr G\cr\alpha\cr\beta\cr\mathcal{F}\cr\mathcal{G}\cr}
\nonumber\\
   &&\qquad+a^4\sum_x
   ((\chi P_+)^T\,\,(\chi P_-)^T\,\,(XP_+)^T\,\,(XP_-)^T)
   M_{\rm F}
   \pmatrix{P_+\chi\cr P_-\chi\cr P_+X\cr P_-X\cr},
\end{eqnarray}
where matrices $M_{\rm B}$ and~$M_{\rm F}$ are given by
\begin{eqnarray}
   &&M_{\rm B}
\nonumber\\
   &&=
   {1\over2}\pmatrix{
   D_1^2&0&D_2+R&I&0&0&R&I\cr
   0&D_1^2&-I&D_2+R&0&0&-I&R\cr
   D_2+R&-I&1&0&R&-I&0&0\cr
   I&D_2+R&0&1&I&R&0&0\cr
   0&0&R&I&0&0&-{2\over a}+R&I\cr
   0&0&-I&R&0&0&-I&-{2\over a}+R\cr
   R&-I&0&0&-{2\over a}+R&-I&0&0\cr
   I&R&0&0&I&-{2\over a}+R&0&0\cr},
\nonumber\\
\end{eqnarray}
where $R=\re m$ and $I=\im m$ and
\begin{equation}
   M_{\rm F}=
   {1\over2}C\pmatrix{
   D_2+m&D_1&m&0\cr
   D_1&D_2+m^*&0&m^*\cr
   m&0&-{2\over a}+m&0\cr
   0&m^*&0&-{2\over a}+m^*\cr}.
\end{equation}
By noting the relation~(\ref{twoxeleven}), we then find that gaussian
integrations over bosonic variables give rise to the following contribution to
the partition function
\begin{equation}
   \det\nolimits^{-1}\Bigl\{{-1\over4a^2}
   \Bigl[{2\over a}D_2+(1-{1\over2}aD_2)m^*m\Bigr]\Bigr\}.
\end{equation}
On the other hand, integrations over fermionic variables give rise to
\begin{equation}
   \det C\det\Bigl\{{1\over4a}
   \Bigl[{2\over a}D_2+(1-{1\over2}aD_2)m^*m\Bigr]\Bigr\},
\end{equation}
where we have taken into account the fact that $\chi$ and~$X$ are 4~component
spinors. These contributions from bosonic and fermionic variables are cancelled
to each other, leaving a constant which may be normalized to unity. Therefore,
the cosmological term is not generated by one-loop radiative corrections
{\it even for}~$a\neq0$, as expected from the exact supersymmetry of the free
action~$S_0$.

\listoftables           
\listoffigures          

\end{document}